\documentclass[longauth]{aa}
\usepackage{bm}
\usepackage{natbib}
\bibpunct{(}{)}{;}{a}{}{,} 
\usepackage{graphicx}
\usepackage{txfonts}
\usepackage[colorlinks=true, allcolors=blue]{hyperref}
\usepackage{bm}
\usepackage{multirow}
\usepackage{euclid}
\usepackage{changepage}
\usepackage{booktabs}
\defcitealias{EuclidCollaboration2019}{EC20}
\defcitealias{Collaboration2016}{Planck Collaboration 2016}

\hypersetup{
    colorlinks=true,
    linkcolor=blue,
    citecolor=blue,
}

\newcommand{\bv}{b_\mathrm{v}}
\newcommand{\bveff}{b_{\mathrm{v}}^\mathrm{eff}}
\newcommand{\bveffi}{b_{\mathrm{v},\,i}^\mathrm{eff}}
\newcommand{\nivz}{n_i^\textrm{v}(z)}
\newcommand{\vis}{VIS}
\newcommand{\nisp}{NISP}

\newcommand{\pmkz}{P_{\mathrm{mm}}(k,z)}
\newcommand{\pmlimb}{P_{\textrm{mm}}\left[\frac{\ell+1 / 2}{r(z)}, z\right]}
\newcommand{\cl}{C(\ell)}
\newcommand{\clest}{\hat{C}(\ell)}
\newcommand{\clgg}{C^{\gamma\gamma}(\ell)}
\newcommand{\clvv}{C^\textrm{vv}(\ell)}
\newcommand{\clvg}{C^{\textrm{v}\gamma}(\ell)}
\newcommand{\clABij}{C^\mathrm{AB}_{ij}(\ell)}
\newcommand{\clestABij}{\hat{C}^{\mathrm{AB}}_{ij}(\ell)}
\newcommand{\clggij}{C^{\gamma \gamma}_{ij}(\ell)}
\newcommand{\clvvij}{C^\textrm{vv}_{ij}(\ell)}
\newcommand{\clgvij}{C^{\textrm{v}\gamma}_{ij}(\ell)}

\newcommand{\fisher}{F_{\alpha \beta}}
\newcommand{\camb}{\texttt{CAMB}}
\newcommand{\lcdm}{\Lambda \mathrm{CDM}}
\newcommand{\nlcdm}{\nu \lcdm}
\newcommand{\wcdm}{w_0 w_a \mathrm{CDM}}
\newcommand{\nwcdm}{\nu \wcdm}
\newcommand{\Hzero}{H_0}
\newcommand{\HzeroUNIT}{\text{km} \, \text{s}^{-1} \, \text{Mpc}^{-1}}

\newcommand{\Hz}{H(z)}
\newcommand{\Omm}{\Omega_\textrm{m}}

\newcommand{\Omb}{\Omega_\textrm{b}}

\newcommand{\Omc}{\Omega_\textrm{c}}

\newcommand{\Omn}{\Omega_\nu}

\newcommand{\OmL}{\Omega_\Lambda}
\newcommand{\wz}{w_0}
\newcommand{\wa}{w_a}
\newcommand{\Mnu}{M_\nu}
\newcommand{\ns}{n_\textrm{s}}

\newcommand{\sige}{\sigma_8}

\newcommand{\refMnu}{0.06}

\newcommand{\wklens}{\mathrm{WL}}

\newcommand{\Voids}{\mathrm{V}}
\newcommand{\xc}{\mathrm{WL+V+XC}}
\newcommand{\wlpv}{\mathrm{WL+V}}

\newcommand{\deltac}{\delta_\mathrm{c}}
\newcommand{\deltav}{\delta_\mathrm{v}}
\newcommand{\pzpz}{p_{\mathrm{ph}}\left(z_{\mathrm{p}} | z\right)}
\newcommand{\ngz}{n^\mathrm{g}(z)}
\newcommand{\nigz}{n_i^\mathrm{g}(z)}
\renewcommand\aap{A\&A}

\renewcommand\apjl{ApJL}

\begin{document} 

\title{\Euclid: Forecasts from the void-lensing cross-correlation\thanks{This paper is published on behalf of the Euclid Consortium.}}

\newcommand{\orcid}[1]{} 
\author{M.~Bonici$^{1,2}$\thanks{\email{marco.bonici@inaf.it}}, C.~Carbone$^{2}$, S.~Davini$^{3}$, P.~Vielzeuf\orcid{0000-0003-2035-9339}$^{4}$, L.~Paganin$^{1,3}$, V.~Cardone$^{5,6}$, N.~Hamaus\orcid{0000-0002-0876-2101}$^{7}$, A.~Pisani\orcid{0000-0002-6146-4437}$^{8}$, A.J.~Hawken$^{4}$, A.~Kovacs\orcid{0000-0002-5825-579X}$^{9,10}$
% ...
}
\institute{$^{1}$ Dipartimento di Fisica, Universit\'a degli studi di Genova, and INFN-Sezione di Genova, via Dodecaneso 33, I-16146, Genova, Italy\\
$^{2}$ INAF-IASF Milano, Via Alfonso Corti 12, I-20133 Milano, Italy\\
$^{3}$ INFN-Sezione di Genova, Via Dodecaneso 33, I-16146, Genova, Italy\\
$^{4}$ Aix-Marseille Univ, CNRS/IN2P3, CPPM, Marseille, France\\
$^{5}$ INAF-Osservatorio Astronomico di Roma, Via Frascati 33, I-00078 Monteporzio Catone, Italy\\
$^{6}$ INFN-Sezione di Roma, Piazzale Aldo Moro, 2 - c/o Dipartimento di Fisica, Edificio G. Marconi, I-00185 Roma, Italy\\
$^{7}$ Universit\"ats-Sternwarte M\"unchen, Fakult\"at f\"ur Physik, Ludwig-Maximilians-Universit\"at M\"unchen, Scheinerstrasse 1, 81679 M\"unchen, Germany\\
$^{8}$ Department of Astrophysical Sciences, Peyton Hall, Princeton University, Princeton, NJ 08544, USA\\
$^{9}$ Instituto de Astrof\'isica de Canarias (IAC); Departamento de Astrof\'isica, Universidad de La Laguna (ULL), E-38200, La Laguna, Tenerife, Spain\\
$^{10}$ Instituto de Astrof\'isica de Canarias, Calle V\'ia L\'actea s/n, E-38204, San Crist\'obal de La Laguna, Tenerife, Spain\\
$^{11}$ Department of Physics and Astronomy, University College London, Gower Street, London WC1E 6BT, UK\\
$^{12}$ Dipartimento di Fisica e Astronomia "Augusto Righi" - Alma Mater Studiorum Universit\`{a} di Bologna, via Piero Gobetti 93/2, I-40129 Bologna, Italy\\
$^{13}$ INAF-Osservatorio di Astrofisica e Scienza dello Spazio di Bologna, Via Piero Gobetti 93/3, I-40129 Bologna, Italy\\
$^{14}$ INFN-Sezione di Bologna, Viale Berti Pichat 6/2, I-40127 Bologna, Italy\\
$^{15}$ Dipartimento di Fisica e Astronomia "G.Galilei", Universit\'a di Padova, Via Marzolo 8, I-35131 Padova, Italy\\
$^{16}$ INFN-Padova, Via Marzolo 8, I-35131 Padova, Italy\\
$^{17}$ Universit\'e de Gen\`eve, D\'epartement de Physique Th\'eorique and Centre for Astroparticle Physics, 24 quai Ernest-Ansermet, CH-1211 Gen\`eve 4, Switzerland\\
$^{18}$ Institute of Space Sciences (ICE, CSIC), Campus UAB, Carrer de Can Magrans, s/n, 08193 Barcelona, Spain\\
$^{19}$ Institut d'Estudis Espacials de Catalunya (IEEC), Carrer Gran Capit\'a 2-4, 08034 Barcelona, Spain\\
$^{20}$ Institut de Recherche en Astrophysique et Plan\'etologie (IRAP), Universit\'e de Toulouse, CNRS, UPS, CNES, 14 Av. Edouard Belin, F-31400 Toulouse, France\\
$^{21}$ University of Lyon, UCB Lyon 1, CNRS/IN2P3, IUF, IP2I Lyon, France\\
$^{22}$ Istituto Nazionale di Astrofisica (INAF) - Osservatorio di Astrofisica e Scienza dello Spazio (OAS), Via Gobetti 93/3, I-40127 Bologna, Italy\\
$^{23}$ Istituto Nazionale di Fisica Nucleare, Sezione di Bologna, Via Irnerio 46, I-40126 Bologna, Italy\\
$^{24}$ Institute for Astronomy, University of Edinburgh, Royal Observatory, Blackford Hill, Edinburgh EH9 3HJ, UK\\
$^{25}$ Dipartimento di Fisica, Universit\'a degli Studi di Torino, Via P. Giuria 1, I-10125 Torino, Italy\\
$^{26}$ INFN-Sezione di Torino, Via P. Giuria 1, I-10125 Torino, Italy\\
$^{27}$ INAF-Osservatorio Astrofisico di Torino, Via Osservatorio 20, I-10025 Pino Torinese (TO), Italy\\
$^{28}$ Instituto de F\'isica Te\'orica UAM-CSIC, Campus de Cantoblanco, E-28049 Madrid, Spain\\
$^{29}$ AIM, CEA, CNRS, Universit\'{e} Paris-Saclay, Universit\'{e} de Paris, F-91191 Gif-sur-Yvette, France\\
$^{30}$ Universit\'e Paris-Saclay, Universit\'e Paris Cit\'e, CEA, CNRS, Astrophysique, Instrumentation et Mod\'elisation Paris-Saclay, 91191 Gif-sur-Yvette, France\\
$^{31}$ Universit\'e St Joseph; Faculty of Sciences, Beirut, Lebanon\\
$^{32}$ Departamento de F\'isica, FCFM, Universidad de Chile, Blanco Encalada 2008, Santiago, Chile\\
$^{33}$ Astrophysics Research Institute, Liverpool John Moores University, 146 Brownlow Hill, Liverpool L3 5RF, UK\\
$^{34}$ Institute of Cosmology and Gravitation, University of Portsmouth, Portsmouth PO1 3FX, UK\\
$^{35}$ Dipartimento di Fisica e Astronomia, Universit\'a di Bologna, Via Gobetti 93/2, I-40129 Bologna, Italy\\
$^{36}$ Dipartimento di Fisica, Universit\`{a} di Genova, Via Dodecaneso 33, I-16146, Genova, Italy\\
$^{37}$ INFN-Sezione di Roma Tre, Via della Vasca Navale 84, I-00146, Roma, Italy\\
$^{38}$ INAF-Osservatorio Astronomico di Capodimonte, Via Moiariello 16, I-80131 Napoli, Italy\\
$^{39}$ Instituto de Astrof\'isica e Ci\^encias do Espa\c{c}o, Universidade do Porto, CAUP, Rua das Estrelas, PT4150-762 Porto, Portugal\\
$^{40}$ Institut de F\'{i}sica d'Altes Energies (IFAE), The Barcelona Institute of Science and Technology, Campus UAB, 08193 Bellaterra (Barcelona), Spain\\
$^{41}$ Port d'Informaci\'{o} Cient\'{i}fica, Campus UAB, C. Albareda s/n, 08193 Bellaterra (Barcelona), Spain\\
$^{42}$ INFN section of Naples, Via Cinthia 6, I-80126, Napoli, Italy\\
$^{43}$ Department of Physics "E. Pancini", University Federico II, Via Cinthia 6, I-80126, Napoli, Italy\\
$^{44}$ Centre National d'Etudes Spatiales, Toulouse, France\\
$^{45}$ Institut national de physique nucl\'eaire et de physique des particules, 3 rue Michel-Ange, 75794 Paris C\'edex 16, France\\
$^{46}$ European Space Agency/ESRIN, Largo Galileo Galilei 1, 00044 Frascati, Roma, Italy\\
$^{47}$ ESAC/ESA, Camino Bajo del Castillo, s/n., Urb. Villafranca del Castillo, 28692 Villanueva de la Ca\~nada, Madrid, Spain\\
$^{48}$ Univ Lyon, Univ Claude Bernard Lyon 1, CNRS/IN2P3, IP2I Lyon, UMR 5822, F-69622, Villeurbanne, France\\
$^{49}$ Institute of Physics, Laboratory of Astrophysics, Ecole Polytechnique F\'{e}d\'{e}rale de Lausanne (EPFL), Observatoire de Sauverny, 1290 Versoix, Switzerland\\
$^{50}$ Mullard Space Science Laboratory, University College London, Holmbury St Mary, Dorking, Surrey RH5 6NT, UK\\
$^{51}$ Departamento de F\'isica, Faculdade de Ci\^encias, Universidade de Lisboa, Edif\'icio C8, Campo Grande, PT1749-016 Lisboa, Portugal\\
$^{52}$ Instituto de Astrof\'isica e Ci\^encias do Espa\c{c}o, Faculdade de Ci\^encias, Universidade de Lisboa, Campo Grande, PT-1749-016 Lisboa, Portugal\\
$^{53}$ Department of Astronomy, University of Geneva, ch. d\'Ecogia 16, CH-1290 Versoix, Switzerland\\
$^{54}$ Universit\'e Paris-Saclay, CNRS, Institut d'astrophysique spatiale, 91405, Orsay, France\\
$^{55}$ Department of Physics, Oxford University, Keble Road, Oxford OX1 3RH, UK\\
$^{56}$ INAF-Osservatorio Astronomico di Trieste, Via G. B. Tiepolo 11, I-34143 Trieste, Italy\\
$^{57}$ FRACTAL S.L.N.E., calle Tulip\'an 2, Portal 13 1A, 28231, Las Rozas de Madrid, Spain\\
$^{58}$ INAF-Osservatorio Astronomico di Padova, Via dell'Osservatorio 5, I-35122 Padova, Italy\\
$^{59}$ Max Planck Institute for Extraterrestrial Physics, Giessenbachstr. 1, D-85748 Garching, Germany\\
$^{60}$ Dipartimento di Fisica "Aldo Pontremoli", Universit\'a degli Studi di Milano, Via Celoria 16, I-20133 Milano, Italy\\
$^{61}$ INAF-Osservatorio Astronomico di Brera, Via Brera 28, I-20122 Milano, Italy\\
$^{62}$ INFN-Sezione di Milano, Via Celoria 16, I-20133 Milano, Italy\\
$^{63}$ Institute of Theoretical Astrophysics, University of Oslo, P.O. Box 1029 Blindern, N-0315 Oslo, Norway\\
$^{64}$ Jet Propulsion Laboratory, California Institute of Technology, 4800 Oak Grove Drive, Pasadena, CA, 91109, USA\\
$^{65}$ von Hoerner \& Sulger GmbH, Schlo{\ss}Platz 8, D-68723 Schwetzingen, Germany\\
$^{66}$ Technical University of Denmark, Elektrovej 327, 2800 Kgs. Lyngby, Denmark\\
$^{67}$ Max-Planck-Institut f\"ur Astronomie, K\"onigstuhl 17, D-69117 Heidelberg, Germany\\
$^{68}$ Department of Physics and Helsinki Institute of Physics, Gustaf H\"allstr\"omin katu 2, 00014 University of Helsinki, Finland\\
$^{69}$ European Space Agency/ESTEC, Keplerlaan 1, 2201 AZ Noordwijk, The Netherlands\\
$^{70}$ NOVA optical infrared instrumentation group at ASTRON, Oude Hoogeveensedijk 4, 7991PD, Dwingeloo, The Netherlands\\
$^{71}$ Argelander-Institut f\"ur Astronomie, Universit\"at Bonn, Auf dem H\"ugel 71, 53121 Bonn, Germany\\
$^{72}$ Department of Physics, Institute for Computational Cosmology, Durham University, South Road, DH1 3LE, UK\\
$^{73}$ University of Applied Sciences and Arts of Northwestern Switzerland, School of Engineering, 5210 Windisch, Switzerland\\
$^{74}$ INFN-Bologna, Via Irnerio 46, I-40126 Bologna, Italy\\
$^{75}$ Department of Physics and Astronomy, University of Aarhus, Ny Munkegade 120, DK-8000 Aarhus C, Denmark\\
$^{76}$ Centre for Astrophysics, University of Waterloo, Waterloo, Ontario N2L 3G1, Canada\\
$^{77}$ Department of Physics and Astronomy, University of Waterloo, Waterloo, Ontario N2L 3G1, Canada\\
$^{78}$ Perimeter Institute for Theoretical Physics, Waterloo, Ontario N2L 2Y5, Canada\\
$^{79}$ Space Science Data Center, Italian Space Agency, via del Politecnico snc, 00133 Roma, Italy\\
$^{80}$ Institute of Space Science, Bucharest, Ro-077125, Romania\\
$^{81}$ Departamento de Astrof\'{i}sica, Universidad de La Laguna, E-38206, La Laguna, Tenerife, Spain\\
$^{82}$ IFPU, Institute for Fundamental Physics of the Universe, via Beirut 2, 34151 Trieste, Italy\\
$^{83}$ Aix-Marseille Univ, CNRS, CNES, LAM, Marseille, France\\
$^{84}$ Centro de Investigaciones Energ\'eticas, Medioambientales y Tecnol\'ogicas (CIEMAT), Avenida Complutense 40, 28040 Madrid, Spain\\
$^{85}$ Instituto de Astrof\'isica e Ci\^encias do Espa\c{c}o, Faculdade de Ci\^encias, Universidade de Lisboa, Tapada da Ajuda, PT-1349-018 Lisboa, Portugal\\
$^{86}$ Universidad Polit\'ecnica de Cartagena, Departamento de Electr\'onica y Tecnolog\'ia de Computadoras, 30202 Cartagena, Spain\\
$^{87}$ Kapteyn Astronomical Institute, University of Groningen, PO Box 800, 9700 AV Groningen, The Netherlands\\
$^{88}$ Infrared Processing and Analysis Center, California Institute of Technology, Pasadena, CA 91125, USA}
\date{Received ; accepted }
\abstract{
The \Euclid space telescope will survey a large dataset of cosmic voids traced by dense samples of galaxies. In this work we estimate its expected performance when exploiting angular photometric void clustering, galaxy weak lensing, and their cross-correlation. To this aim, we implemented a Fisher matrix approach tailored for voids from the \Euclid photometric dataset and we present the first forecasts on cosmological parameters that include the void-lensing correlation. We examined two different probe settings, pessimistic and optimistic, both for void clustering and galaxy lensing.
We carried out forecast analyses in four model cosmologies, accounting for a varying total neutrino mass, $\Mnu$, and a dynamical dark energy (DE) equation of state, $w(z)$, described by the popular Chevallier-Polarski-Linder parametrization. We find that void clustering constraints on $h$ and $\Omb$ are competitive with galaxy lensing alone, while errors on $\ns$ decrease thanks to the orthogonality of the two probes in the 2D-projected parameter space.
We also note that, as a whole, with respect to assuming the two probes as independent, the inclusion of the void-lensing cross-correlation signal improves parameter constraints by $10-15\%$, and enhances the joint void clustering and galaxy lensing figure of merit (FoM) by $10\%$ and $25\%$, in the pessimistic and optimistic scenarios, respectively. Finally, when further combining with the spectroscopic galaxy clustering, assumed as an independent probe, we find that, in the most competitive case, the FoM increases by a factor of 4 with respect to the combination of weak lensing and spectroscopic galaxy clustering taken as independent probes. The forecasts presented in this work show that photometric void clustering and its cross-correlation with galaxy lensing deserve to be exploited in the data analysis of the \Euclid galaxy survey and promise to improve its constraining power, especially on $h$, $\Omb$, the neutrino mass, and the DE evolution.}

\date{}

\authorrunning{M. Bonici et al.}

\titlerunning{\Euclid: Forecasts from the void-lensing cross-correlation}

\maketitle

\section{Introduction}
\label{sec:introduction}

The late-time cosmic acceleration of the Universe on large scales is an established observational fact~\citep{Riess1998, Perlmutter1998}. The cosmological standard model, the so-called $\lcdm$ model, explains the acceleration as the effect of a cosmological constant $\Lambda$, postulating the existence of a dark energy (DE) component with negative pressure, whose physical nature remains poorly understood. Minimal extensions to such a model include a variation of DE with time, parametrized by an equation of state following Chevallier-Polarski-Linder (CPL): $w(z) = w_0 + z/(1+z)\, \wa$ \citep{chevallier2001accelerating, Linder2002}.
Such pressure impacts not only the background evolution of the Universe, but also the growth of large-scale structure (LSS): the greater the acceleration, the slower the structures can grow.
While different DE models have been proposed~\citep{Ratra1988, Steinhardt1998, 2000PhRvL..85.4438A, Copeland2006}, the physical nature of DE remains unknown. Precise measurements of $w(z)$ promise to shed light on its features.

Massive relic neutrinos can also affect the growth of structure  and the expansion history of the Universe. Due to their free streaming with hot thermal velocities~\citep{Lesgourgues2006, Kiakotou2007, Lesgourgues2012, Lesgourgues2014},  they alter the epoch of matter-radiation equality and suppress the growth of structure  at mildly nonlinear and small scales.
The Universe can thus be used as a laboratory to constrain the neutrino mass scale.

Upcoming galaxy surveys are designed to probe the DE equation of state $w$ and the neutrino mass scale, with direct observations and measurements of the growth of structure and of distance scales.
Cosmological probes used in these investigations include galaxy clustering (GC) and weak lensing (WL).

At the GC level, measurements of the galaxy correlation function from LSS surveys are used to constrain cosmological parameters, exploiting the sensitivity of the galaxy density fluctuations to the underlying dark matter (DM) density field. At the WL level, images of large ensembles of galaxies provide
the so-called \textit{cosmic shear}: tiny distortions in the shapes of galaxies due to the gravitational potential produced by intervening density perturbations, crossed by light propagating from the source to the observer.
Images of galaxy shapes, complemented by an estimation of their redshifts, allow us to measure the structure growth and improve the inference on cosmological parameters.

Recently, aside from traditional GC, under-dense regions known as cosmic voids have been used to extract cosmological information \citep{Lavaux2010, Sutter2014, Hamaus2017, Nadathur2019, Pisani2019, Hamaus2016, Hamaus2020}. Their sizes range from about ten to a few hundreds of $\text{Mpc}$~\citep{Sheth2004}.
Galaxy redshift surveys allow us to build cosmic void catalogs \citep{Sutter2012, Micheletti2014, Clampitt2015, Sanchez2017,  Mao2017,Hawken2019, Pollina2019, Hamaus2020,  Aubert2020, Nadathur2020}. As for galaxies and galaxy clusters, it is possible to study the correlation function of cosmic voids and their possible cross-correlation with other cosmological probes~\citep{Granett2008, Ilic2013, Hamaus2014a, Granett2015}.
In particular, it is possible to measure the cross-correlation between WL and under-dense regions, and exploit this signal to infer cosmological information, and break possible parameter degeneracy \citep{Krause2013, Melchior2014, Clampitt2015, Sanchez2017, Fang2019a}.

The use of cosmic voids as a cosmological probe presents several advantages. Firstly, voids are less affected than DM halos by shell-crossing and virialization effects, which makes their dynamical evolution more amenable to theoretical models and easier to describe~\citep{Sheth2004, Hamaus2014b, Stopyra2020}.

Secondly, recent works in the literature~\citep[e.g.,][]{Lavaux2012,Pisani2015, Massara2015a, Kreisch2019, Schuster2019, Verza2019b, Pisani2019, Bos2012, Lee2009, Kreisch2021} have shown that void formation and evolution are sensitive to the DE equation of state, and to the total neutrino mass, in a qualitatively and quantitatively different way with respect to over-dense structures, because of the different scales involved and their different nature. 

For instance, the DE contribution within voids can be more dominant than in the Universe, on average, allowing such a probe to have a large sensitivity to its amount and evolution. In this work we consider DE as an effective quintessence “fluid” that does not cluster significantly and is nearly homogeneously distributed in space. Due to their hot thermal velocities, massive neutrinos have a free-streaming length that, depending on their mass and redshift, can range between one hundred to a few tens of $\hMpc$~\citep{Lesgourgues2006}, thus matching the typical void sizes~\citep{Kreisch2019}.

This paper presents the first forecasts on cosmological parameters based on the combination of WL, void-lensing cross-correlation, and angular void clustering, where voids are found in the galaxy photometric catalog of the \Euclid survey, an ESA medium-class mission, currently scheduled
for launch in 2023. The paper belongs to a series of companion papers investigating the scientific return that can be expected from voids from the Euclid mission~\citep{Hamaus2021a, Contarini2022}.

It is expected that the \Euclid photometric galaxy catalog will be characterized by a surface density of  $n_\mathrm{g} = 30\, {\rm arcmin}^{-2}$~\citep{laureijs2011euclid}, and the imaging \Euclid catalog will contain the shapes of about 1.5 billion galaxies, observed in the visible range with the \vis\ instrument~\citep{VISCropper2018}. The redshifts of such galaxies will be measured in photometric mode, using the Near Infrared Spectroscopic Photometric (\nisp) instrument~\citep{NISPCostille2018}, complemented by ground-based observations in different bands.

The parameter forecasts presented here closely follow the recipe of the \Euclid Inter Science Task-Force for Forecasts (IST:F) group \citep[][hereafter EC20]{EuclidCollaboration2019} in the case of WL and photometric galaxy clustering (${\rm GC}_{\rm ph}$): the covariance between cosmological parameters is evaluated with the Fisher matrix approach, using angular power spectra computed within the Limber approximation~\citep{Limber1953}, and exploiting a tomography technique.
Forecasts are given for spatially flat $\lcdm$, $\nlcdm$, $\wcdm$, and $\nwcdm$ model cosmologies, adopting the following set of cosmological parameters: the reduced Hubble constant $h$ defined via $H_0\,=\,h\,100\,\HzeroUNIT$, the baryon density parameter $\Omb$ at present time, the total matter density parameter $\Omm$ at present time, the sum of the three active neutrino masses $\Mnu=\sum_i m_{\nu_i}$, the massive neutrino density parameter $\Omn=\Mnu \, ({\rm eV})/(93.14 \,h^2)$ at present time, the cold DM density parameter $\Omc=\Omm-\Omb-\Omn$ at present time, where $\Omm = \Omb +\Omc +\Omn$, the DE density parameter $\Omega_{\rm DE}= 1-\Omm$ at present time, the parameters of the DE equation of state, $\wz$ and $\wa$, the r.m.s of the matter linear density fluctuations,  $\sige$, inside a radius of $8\,\hMpc$, and the scalar spectral index $\ns$.
The cosmological parameters varied in the analysis, together with their fiducial values, are summarized in Table~\ref{table:par_ref}.

\begin{table*}
        \caption{Cosmological parameters varied in the forecast analysis, together with their values in the fiducial cosmology assumed in this work.}
        \centering
        \begin{tabular}{ c| c |c |c |c |c |c |c }
        \toprule
        $\Omb$ & $\Omm$ & $\wz$ & $\wa$ & $h$ & $\Mnu$ [eV] & $\ns$ & $\sige$ \\
                \midrule
        0.05 & 0.32 & $-1$ & 0 & 0.67 & \refMnu & 0.96 & 0.816 \\  
                \bottomrule     
        \end{tabular}
\label{table:par_ref}
\end{table*}

The article is organized as follows.
Section~\ref{sec:void} illustrates the theoretical modeling of void clustering and bias.
Section~\ref{sec:cl} describes the angular power spectra exploited as observables for parameter forecasts.
Section~\ref{sec:fisher} details the evaluation of the Fisher matrix.
Section~\ref{sec:results} reports the results and interpretation of our analysis.
Finally, Sect.~\ref{sec:conclusion} presents our concluding remarks. 

\section{The clustering of cosmic voids}
\label{sec:void}

\begin{figure*}[tbp]
\hspace*{-0.4cm}
\begin{tabular}{c@{}c}
        \includegraphics[width=.505\textwidth]{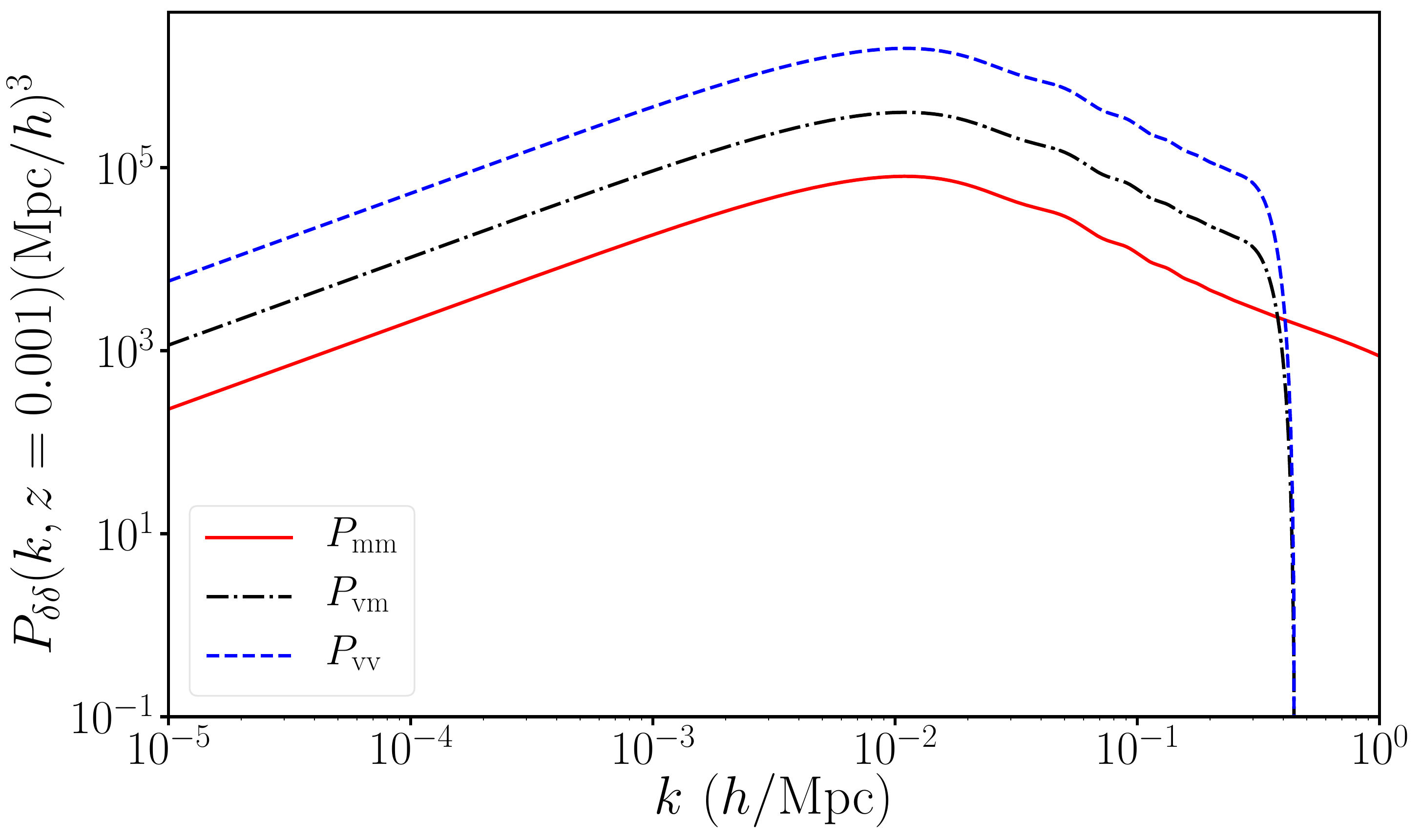} &      
    \includegraphics[width=0.505\textwidth]{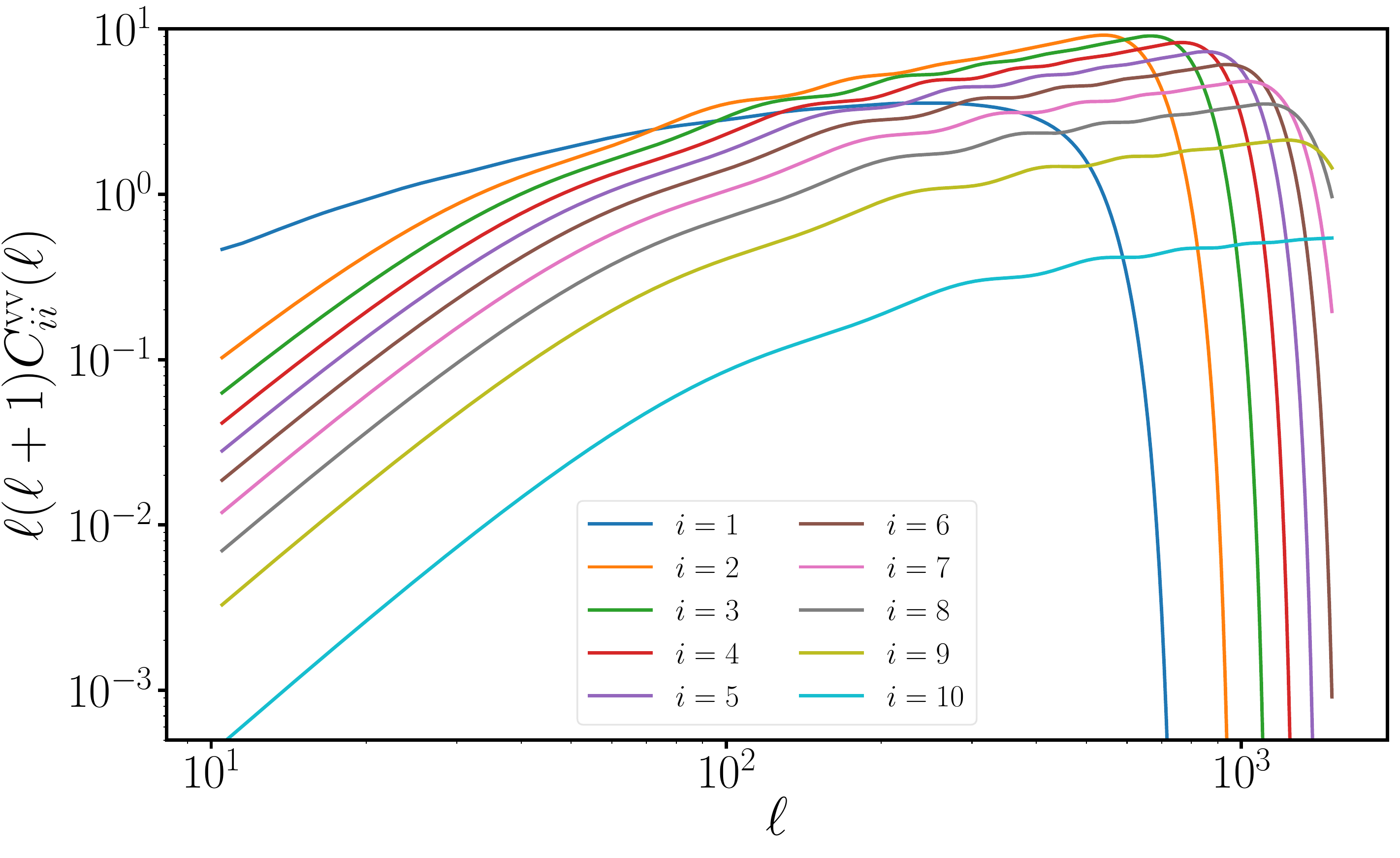}\\
             \includegraphics[width=0.505\textwidth]{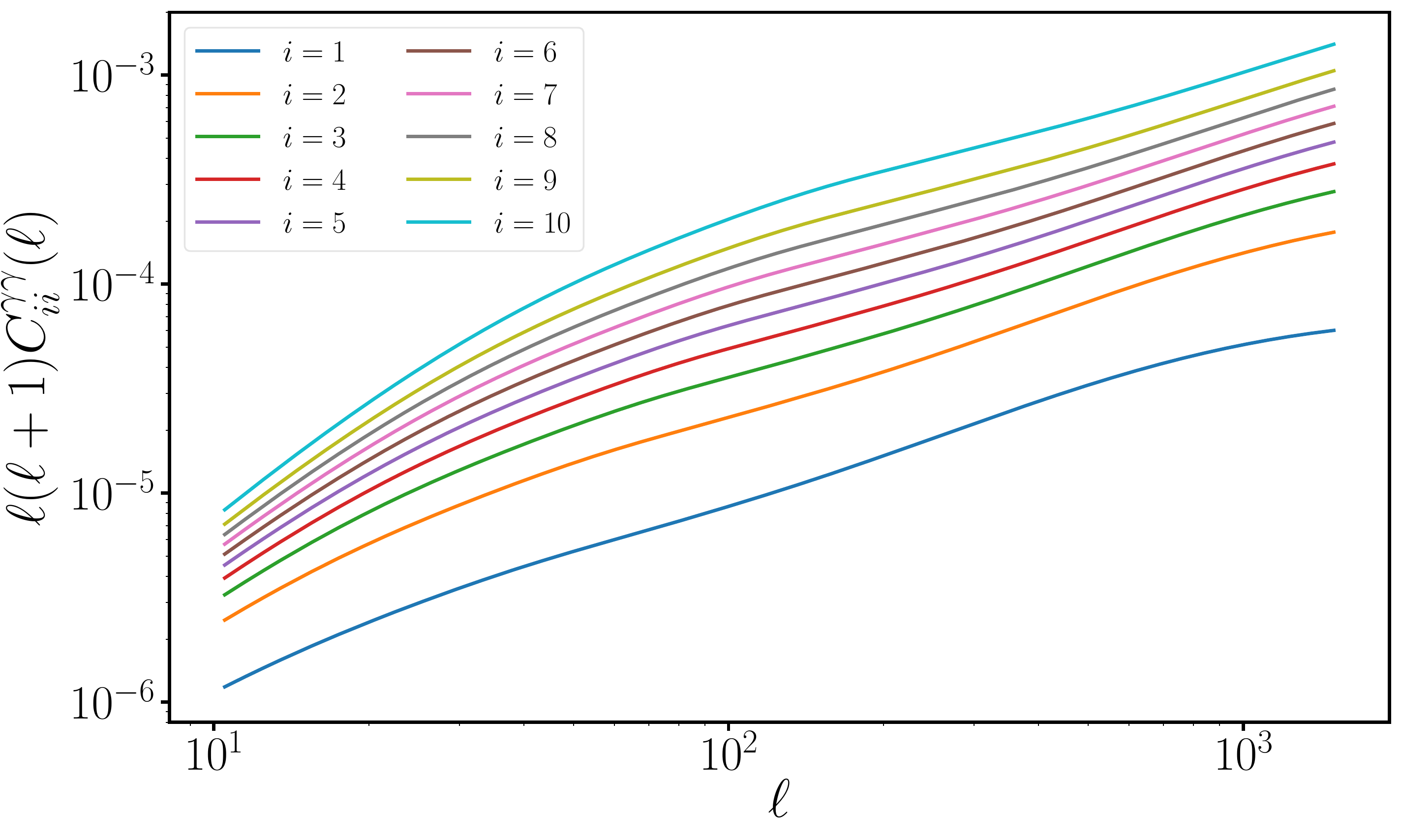}&
     \includegraphics[width=0.505\textwidth]{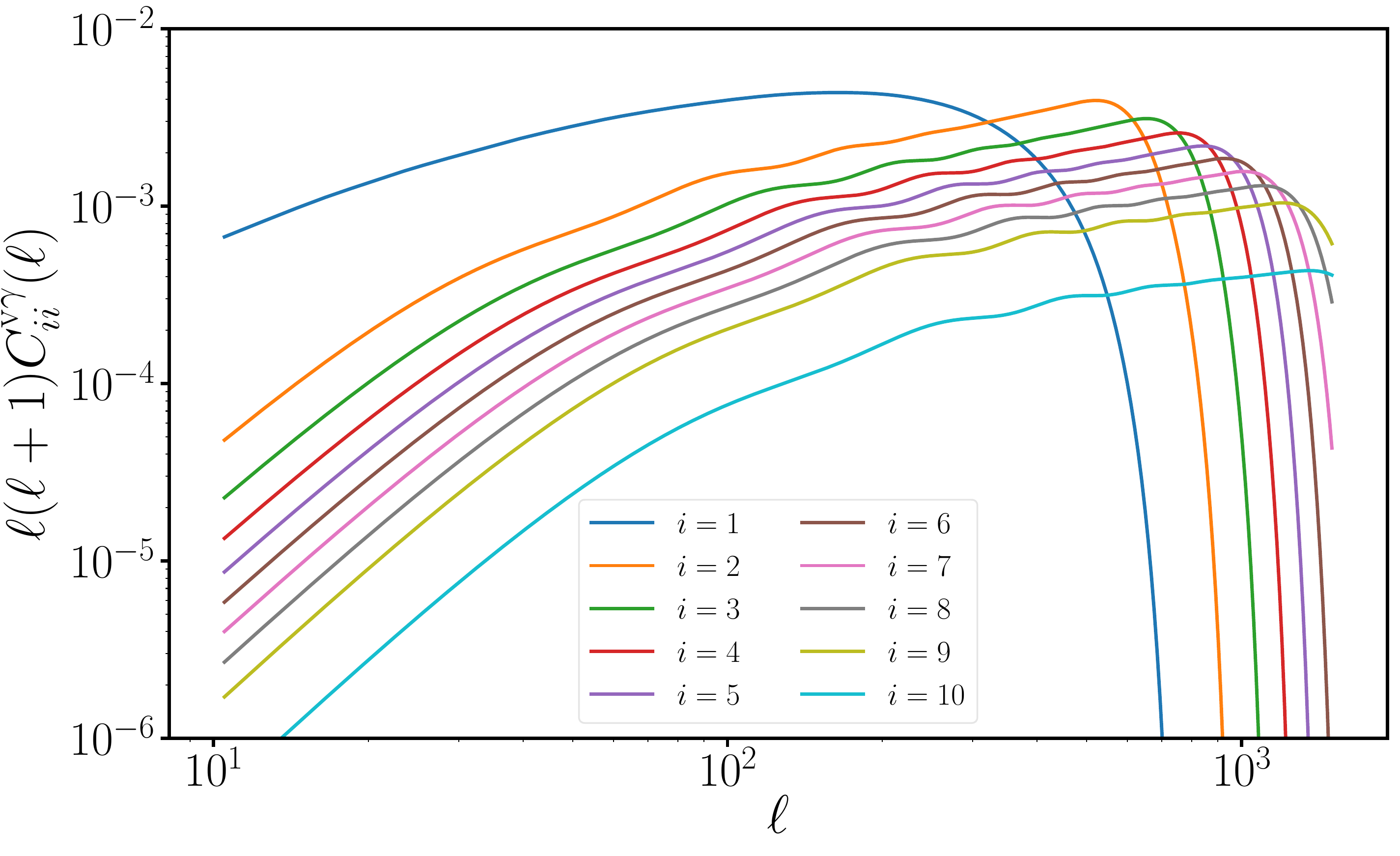}
\end{tabular}
        \caption{Figure containing the power spectra used in this work. {\it Top left}: Nonlinear matter auto-power spectrum $P_\textrm{mm}$ (solid red line), the void auto-power spectrum $P_\textrm{vv}$ without the shot noise (dashed blue line), computed assuming the pessimistic void bias scenario with $b_{v\mathrm{eff}}\approx -11.9$, and the absolute value of the void-matter cross-power spectrum $P_\textrm{vm}$ (dot-dashed black line), all computed at redshift $z=0.001$.
{\it Top right}: Tomographic void angular auto-power spectra $\clvvij$ for the diagonal ($i=j$) tomographic bins, without shot noise. {\it Bottom left}: Tomographic lensing auto-power spectra $\clggij$ for the diagonal ($i=j$) tomographic bins, without shape noise. {\it Bottom right}: Tomographic void-lensing angular cross-spectra $\clgvij$ for the diagonal ($i=j$) tomographic bins. All the spectra are theoretically evaluated in the reference cosmology reported in Table~\ref{table:par_ref} and the bin endpoints are given in Eq.~\eqref{eq:tomo_bins}.}
        \label{fig:cl_plots}
\end{figure*}

We evaluated the angular power spectrum, $C_{\ell}^{\mathrm{vv}}$, of cosmic voids from the void auto-power spectrum, $P_\mathrm{vv} (k, z)$, while the angular cross-power spectra, $C_{\ell}^{\gamma \mathrm{v}}$, between voids and WL were evaluated from the void-matter cross-spectrum $P_{\mathrm{vm}}(k,z)$. The latter are shown, together with the nonlinear matter power spectrum $P_\textrm{mm}(k,z)$, in the top left panel of Fig.~\ref{fig:cl_plots}.  $P_\textrm{vv} (k, z)$ was obtained from the matter power spectrum, $\pmkz$, assuming the Poisson shot noise~\citep{Hamaus2014a, Chan2014, Chan2019, Jamieson2019} and the ``effective void bias'' $\bveff(z)$ defined in Eq.~\ref{eq:bvz}. To this aim we adopted the following relations:
\begin{align}
        \label{eq:bvsq}
        P_\textrm{vv} (k, z) &= \left[ 1-\mathrm{S}_{N}(k) \right] \, \bveff(z)^2
  \, \pmkz \, + 1/\bar{n}^\mathrm{v}(z) \nonumber\\
        &\equiv \bveff(z)^2\, \hat{P}_\mathrm{mm}(k,z)+ 1/\bar{n}^\mathrm{v}(z)\, ,
\end{align}
where the void bias, $\bveff(z)$, is assumed to be scale-independent\footnote{In the presence of massive neutrinos, the void bias could become scale dependent already at the linear level~\citep{Schuster2019}, as is also the case for DM halos and galaxies. However, given the small values of the total neutrino mass considered in this work, we assumed the void bias to be scale independent, following the same approach adopted for galaxies in~\citetalias{EuclidCollaboration2019}.}, $\bar{n}^\mathrm{v}(z)$ is the void number density, and $\hat{P}_\mathrm{mm}(k,z)$ is the nonlinear matter power spectrum when an additional wavenumber filter, $\mathrm{S}_{N}(k)$, is applied to cut small scales.
The low-$k$ pass filter, $\mathrm{S}_{N}(k)$, is necessary since in this work we focus on void clustering at large scales, as, in this case, the behavior of the void power spectrum, $P_\textrm{vv} (k, z)$, is well described by linear theory via a simple multiplicative factor for the void bias. This can be understood in the context of the halo model formalism~\citep{Cooray2002}, which has an analog formulation for voids~\citep{Voivodic2020}. Such an assumption is inaccurate at small scales (excluded from the analyses in the present work), which are sensitive to the void density profile~\citep{Hamaus2014a}. While the inclusion of small scales could improve the results presented in this work, this poses a serious modeling challenge, as  to date in the literature there are no accurate models of the cosmological dependence of the void density profile. Furthermore, we point out that the high shot noise would likely reduce the contribution of the small scales. Therefore, we decided for a more conservative approach, considering only scales independent of the particular void profile, which we can confidently include in our analysis.

Analogously, the void-matter cross-spectrum $P_\textrm{vm}(k,z)$ is defined as
\begin{equation}
        \label{eq:bvmsq}
        P_\textrm{vm} (k, z) = \left[ 1-\mathrm{S}_{N}(k) \right] \, {\bveff(z)}  \, P_\textrm{mm}(k,z) \, \equiv \, \bveff(z)\, \hat{P}_\textrm{mm}(k,z)\, .
\end{equation}

To avoid numerical instabilities, we prefer the low-$k$ pass filter rather than a sharp cut in $k$.  $\mathrm{S}_{N}(k)$ is the so-called smoothstep function, defined as
\begin{equation}
\label{eq:smoothstep}
\mathrm{S}_{N}(k)=\left\{\begin{array}{ll}
0 & \text { if } \frac{k}{k_\mathrm{ref}} \leq a \\
\left(\frac{k}{k_\mathrm{ref}}\right)^{N+1} \sum_{n=0}^{N}\binom{2 N+1}{N-n}\binom{N+n}{n}\left(-\frac{k}{k_\mathrm{ref}}\right)^{n} & \text { if } a \leq \frac{k}{k_\mathrm{ref}} \leq b \\
1 & \text { if } \frac{k}{k_\mathrm{ref}} \geq b\, ,
\end{array}\right.
\end{equation}
where $N$ measures the degree of smoothness of the function itself: the first discontinuous derivative of $\mathrm{S}_{N}(k)$ is the $(N+1)$th derivative.
The order used in $\mathrm{S}_{N}$ for producing $P_\textrm{vv}(k,z)$ and $P_\textrm{vm}(k,z)$ is $N=3$.
The smoothstep parameters are set to:
$a = 1$, $b = 1.8$, and $k_\mathrm{ref}= 0.25h\,\text{Mpc}^{-1}$;
the filter suppresses $P_\textrm{vv}(k,z)$ and $P_\textrm{vm}(k,z)$ for $k>k_{\mathrm{ref}}$. The cutoff scale $k_{\mathrm{ref}}$ was determined combining the mean radius of voids $\bar{r}_\mathrm{v}$ in the catalog with the void exclusion principle, from which we know that the cutoff scale is given by $k\simeq 2\pi/ \bar{r}_\mathrm{v}$ \citep{Hamaus2014a}. We chose this particular filter because it allowed us to better control its effect, since this function has some more parameters. Furthermore, we checked that the exact shape of the filter did not alter sensibly our results, as it was used to remove numerical artifacts.

Summarizing, we define the probe-dependent power spectrum $\hat{P}_\textrm{AB}$ as
\begin{equation}
\label{eq:smoothedpmm}
\hat{P}_\textrm{AB}(k,z)=\left\{\begin{array}{ll}
P_\textrm{mm}(k,z) & \text { if } \textrm{A}=\textrm{B}=\gamma \\
\left[ 1-\mathrm{S}_{N}(k) \right]P_\textrm{mm}(k,z) & \textrm{else}\, ,
\end{array}\right.
\end{equation}
where $\gamma$ stands for cosmic shear.

As we describe below, to obtain the effective void bias, $\bveff(z)$, we used the peak-background split (PBS) formalism~\citep{Sheth2004}, and weighted the void bias, $\bv(z)$, over the void size function~\citep{Sheth2004,Jennings2013}, that is to say, the comoving number density of voids per radius interval.
The void size function was predicted from the ``excursion set formalism'' in the form of the Sheth $\&$ van de Weygart (SvdW) model~\citep{Sheth2004}, according to which voids are treated as isolated objects, whose evolution is, therefore, not affected by the environment and  is assumed to be spherically symmetric.

Under such assumptions, the prediction of the void abundance is completely described by two physical parameters. The former is the density threshold for structure collapse, $\deltac$, linearly extrapolated to the present time as $\deltac(z)=\delta_{\rm c}(0)/G(z)$, where $G(z)$ is the so-called linear growth factor normalized at $z=0$, and $\delta_{\rm c}(0)$ slightly depends on cosmology, here assumed to be $\delta_{\rm c}(0)=1.686$. The latter is the density threshold for the void formation $\deltav$, linearly extrapolated to the present time as $\deltav(z)=\delta_{\rm v}(0)/G(z)$, with $\delta_{\rm v}(0)=-0.9$ \citep[see, e.g.,][recently showing that various values for the threshold $\delta_\mathrm{v}(0)$ can be used to model voids more reliably]{Chan2014, Ronconi2017, Contarini2019, Verza2019b}. We chose a value of $\delta_v(0)$ by comparing our effective bias prediction against the measurement from the Euclid Flagship simulation (more on this in Sect.~\ref{sec:cl}).

In the void abundance characterization, both $\deltav$ and $\deltac$ have an important role. The void formation threshold, $\deltav$, is the value an under-dense region needs to overcome to turn into a void. The density threshold, $\deltac$, is necessary to account for the ``void-in-cloud problem'', where an 
under-dense region inside an over-dense one will not become a void, since the latter will collapse. 

The SvdW model predicts the void abundance to be a function of both $\deltac$ and $\deltav$, via the so-called void-in-cloud parameter, $\mathcal{D}$, and the dimensionless parameter, $x$, defined as
\begin{equation}
\mathcal{D}=\frac{\left|\delta_{\mathrm{v}}\right|}{\delta_{\mathrm{c}}+\left|\delta_{\mathrm{v}}\right|}\, , \quad x=\frac{\mathcal{D}}{\left|\delta_{\mathrm{v}}\right|} \sigma\, ,
\end{equation}
 where $\sigma^2$ is the variance of the filtered linear density field on a scale $R$
\begin{equation}
\sigma^{2}(R) \equiv S(R)=\int \frac{\mathrm{d} k}{k}\,
\frac{k^{3} P_{\mathrm{lin}}(k)}{2\pi^{2}}|W(kR)|^{2}\,.
\end{equation}
Here $P_{\mathrm{lin}}(k)$ is the linear matter power spectrum at $z=0$, and $W(kR)$ is the top-hat filter in Fourier space:
\begin{equation}
W(k R)=\left\{\begin{array}{ll}
1 & kR \leq 1 \\
0 & kR >1\,.
\end{array}\right.
\end{equation}
The abundance of voids with mass $M$ is predicted by the SvdW model as
\begin{equation}
\frac{\mathrm{d} n}{\mathrm{d} \ln M}=\frac{\rho_{\mathrm{m}}}{M} f_{\ln \sigma}(\sigma) \frac{\mathrm{d} \ln \sigma^{-1}}{\mathrm{d} \ln M}\,,
\label{eq:SvdW}
\end{equation}
where $\rho_\mathrm{m}$ is the mean background density of matter and the multiplicity function, $f_{\ln \sigma}$, represents the fraction of voids in a unit range of $\ln M$,
and it is shown to be~\citep{Sheth2004}
\begin{equation}
f_{\ln \sigma}(\sigma)=2 \sum_{j=1}^{\infty} \exp{\left(-\frac{(j \pi x)^{2}}{2}\right)} j \pi x^{2} \sin (j \pi \mathcal{D})\,.
\label{eq:exactmultiplicity}
\end{equation}
Dealing with observations, it is more convenient to work in terms of void volumes rather than void masses.
In linear theory, the relation between the void volume and its mass is
\begin{equation}
\frac{\rho_\mathrm{m}}{M(r_{\mathrm{L}})}= \frac{1}{V(r_{\mathrm{L}})}\,,
\end{equation}
where $V(r_{\mathrm{L}})\equiv 4 \pi r_{\mathrm{L}}^3/3$ is the volume of a sphere with radius $r_{\mathrm{L}}$ computed in the linear regime.

Even if the SvdW approach accurately predicts the void mass abundance, the predicted volume fraction of the Universe inside voids is larger than one. In order to fix this issue, the SvdW approach has been extended to the $V\,{\rm d}n$ size function from \cite{Jennings2013}, who noticed that such a volume fraction issue originates from the incorrect matching between linear and corresponding nonlinear quantities. Since this approach correctly predicts the volume fraction of the Universe inside voids, it is known as the volume-conserving void size function.

\cite{Jennings2013} corrected this match by requiring the void volume fraction of the Universe in the nonlinear regime to have the same value as in the linear one:
\begin{equation}
V(r_{\rm{v}}) \mathrm\, \mathrm{d} n \, = \, \left.V\left(r_\mathrm{L}\right) \mathrm{d} n_{\mathrm{L}}\right|_{r_\mathrm{L}=r_\mathrm{L}(r_{\rm{v}})}\,,
\end{equation}
where the subscript $\rm L$ denotes quantities computed in the linear regime.
The $V\,{\rm d}n$ void size function then reads~\citep{Jennings2013}
\begin{equation}
\frac{\mathrm{d} n}{\mathrm{d} \ln r_{\rm{v}}}=\left.\frac{f_{\ln \sigma}(\sigma)}{V(r_{\rm{v}})} \frac{\mathrm{d} \ln \sigma^{-1}}{\mathrm{d} \ln r_\mathrm{L}} \frac{\mathrm{d} \ln r_\mathrm{L}}{\mathrm{d} \ln r_{\rm{v}}}\right|_{r_\mathrm{L}=r_\mathrm{L}(r_{\rm{v}})}\, ,
\end{equation}
where $r_{\rm{v}}$ indicates the actual void radius.
The conversion from $r_{\mathrm{L}}$ to $r_{\rm{v}}$ is given by
\begin{equation}
\frac{r_{\rm{v}}}{r_{\mathrm{L}}}=\left(\frac{\rho_{\mathrm{m}}}{\rho_{\mathrm{v}}}\right)^{1 / 3}\,,
\end{equation}
where $\rho_\mathrm{v}$ is the average
density within the void.

However, the expression of the multiplicity function, $f_{\ln \sigma}$, in Eq.~\eqref{eq:exactmultiplicity} is difficult to evaluate in practice.
Therefore, \cite{Jennings2013} suggested the following approximation that works at 0.2$\%$ accuracy:
\begin{equation}
f_{\ln \sigma}(\sigma) \approx\left\{\begin{array}{ll}
\sqrt{\frac{2}{\pi}} \frac{\left|\delta_{\mathrm{v}}\right|}{\sigma} \exp{\left(-\frac{\delta_{\mathrm{v}}^{2}}{2 \sigma^{2}}\right)} & x \leqslant 0.276 \\
2 \sum_{j=1}^{4} \exp{\left(-\frac{(j \pi x)^{2}}{2}\right)} j \pi x^{2} \sin (j \pi \mathcal{D}) & x>0.276\,.
\end{array}\right.
\end{equation}

Finally, the $V\,{\rm d}n$ size function is given by
\begin{equation}
\frac{\mathrm{d} n}{\mathrm{d} \ln r_{\rm{v}}}=
\left.\frac{f_{\ln \sigma}(\sigma)}{V(r_{\rm{v}})} 
\frac{\mathrm{d} \ln \sigma^{-1}}{\mathrm{d} \ln r_\mathrm{L}}\right|_{r_\mathrm{L}=r_{\rm{v}}/1.7}\,,
\label{eq:log_Vdn}
\end{equation}
and, after some rearrangements, its final expression is
\begin{equation}
    \frac{\mathrm{d}n}{\mathrm{d}r_{\rm{v}}}=\frac{f_{\ln \sigma}(\sigma)}{r_{\rm{v}} V(r_{\rm{v}})}\left[  -\frac{r_{\mathrm{L}}}{\sigma}\frac{\mathrm{d}\sigma}{\mathrm{d}r_{\mathrm{L}}}\right]_{r_\mathrm{L}=r_{\rm{v}}/1.7}\, ,
    \label{eq:Vdn}
\end{equation}
where $r_{\rm{v}}=1.7 r_{\mathrm{L}} $ is predicted by considering a spherical expansion.

As mentioned previously, the void bias, $\bv$, at a fixed void radius, $r_{\rm{v}}$, is computed using the PBS formalism~\citep{Sheth2004, Chan2014}:
\begin{equation}
\bv (z,r_{\rm{v}}) = 1+\frac{\nu^{2}-1}{\delta_{\mathrm{v}}}+\frac{\delta_{\mathrm{v}} \mathcal{D}}{4 \delta_{\mathrm{c}}^{2} \nu^{2}} \,,
\label{eq:pbs_void_bias}
\end{equation}
where the $z$ dependence is encapsulated in $\delta_{\mathrm{c}}$ and $\delta_{\mathrm{v}}$, and the $r_{\rm{v}}$ dependence in $\nu:$
\begin{equation}
\nu \equiv \frac{|\delta_\mathrm{v}|}{\sigma(r_{\rm{v}})}\, .
\label{eq:bias_modelling}
\end{equation}

We defined the effective void bias by weighting Eq.~\eqref{eq:pbs_void_bias} with the void size function $V\,{\rm d}n$ in Eq.~\eqref{eq:Vdn}:
\begin{equation}
\label{eq:bvz}
\bveff(z) = \frac{\int_{r_{\rm{v},\text
{min}}}^{r_{\rm{v},\text{max}}}\text{d}r_{\rm{v}}\,\frac{\text{d}n}{\text{d}r_{\rm{v}}} \bv(z,r_{\rm{v}} )}{\int_{r_{\rm{v},\text
{min}}}^{r_{\rm{v},\text{max}}}\text{d}r_{\rm{v}}\,\frac{\text{d}n}{\text{d}r_{\rm{v}}}}\, ,
\end{equation}
exploiting in this way all void radii $r \geq r_{\rm{v, min}}\,\hMpc$. In this work, we assumed $r_{\rm{v, min}}=25\,\hMpc$, and we provide forecasts treating the evolution of the void bias in two main scenarios. A ``pessimistic void bias'' scenario, where the redshift evolution of the void bias is supposed to be known from linear theory, but its absolute normalization is unknown. In this case, $\bveff$ is evolved with the growth factor of the fiducial cosmology.
Its fiducial value today is assumed to be $\bveff(z=0) = -11.9$, according to Eq.~\eqref{eq:bvz}, and it is marginalized over as a nuisance parameter. This void bias scenario will be combined with the pessimistic WL scenario as described in Sect.~\ref{sec:cl}. An ``optimistic void bias'' scenario, where we exploit the cosmology dependence of both the void size function and void bias. In particular, such dependence is encapsulated in the growth factor evolution of the density thresholds $\delta_\mathrm{c}$ and $\delta_\mathrm{v}$, which appear both in the void bias definition and in the void size function. This configuration is considered to be optimistic, since we assume that the value of $\bv$, entering $\bveff$, is known from linear theory, and therefore its dependence on the cosmological parameters is exploitable. This void bias scenario will be combined with the optimistic WL scenario as described in Sect.~\ref{sec:cl}.
To summarize, we present forecasts in two scenarios: a pessimistic one, where we treat $\bveff(z=0)$ as an unknown quantity to be marginalized over, fixing its evolution to the fiducial cosmology; and an optimistic one, where we assume we are able to model the cosmological dependence of $\bveff(z)$ and use it as a further source of information on cosmological parameters. Providing forecasts in both cases is, in our opinion, the fairest approach, since we present parameter constraints spanning from the worst to the best case scenario.

\begin{table*}
        \caption{Probe configurations in the pessimistic and optimistic scenarios for the combination V+WL, both considered for two values of $r_{\rm{v, min}}$.}
        \label{tab:configurations}
        \centering
        \begin{tabular}{c | c | c|c|c}
\toprule
        {Configuration}&\multicolumn{2}{c|}{V+WL -- Pessimistic}&\multicolumn{2}{c}{V+WL -- Optimistic}\\
                \midrule
                $\ell$-range (WL) &\multicolumn{2}{c|}{10 -- 1500} & \multicolumn{2}{c}{10 -- 5000}\\ 
                \hline
                        {$\bv$ modeling } &\multicolumn{2}{c|}{$\bveff(z=0)$ nuisance}&\multicolumn{2}{c}{$\bveff(z)$ exploited}\\
                \hline
    $r_{\rm{v, min}}$  & \multicolumn{4}{c}{$25\,\hMpc$}\\

        \hline  
                $\ell$-range (V) &\multicolumn{4}{c}{10 -- 1500}\\ 
                \hline
                \# of evaluated $k$ per decade ($\camb$) &\multicolumn{4}{c}{60}\\ 
                \hline
                \# of evaluated $z$ ($\camb$) &\multicolumn{4}{c}{450}\\
                \hline
                Differentiation method &\multicolumn{4}{c}{SteM}\\
                \bottomrule
        \end{tabular}
\end{table*}

\section{Observables: Angular power spectra}
\label{sec:cl}

This section summarizes the observables used for cosmological parameter forecasts based on WL, voids, and their cross-correlation. Here we describe in particular the  angular power spectra $\cl$, which are functions of the multipole number $\ell$, and are defined as the spherical harmonic transform of a two-point correlation function.
The $a_{\ell m}$, the coefficients of the spherical harmonics decomposition of a field $\mathcal{F}$ on the sphere (e.g., the cosmic shear or void density fields), are defined on the sky as
\begin{equation}
\label{eq:alm_def}
    a_{\ell m} = \int\mathrm{d}\Omega\,\mathcal{F}(\theta,\phi) Y_{\ell m}(\theta,\phi)\,.
\end{equation}

The $\cl$ represent the two-point correlation function of the $a_{\ell m}$ and they are diagonal in $\ell$ and independent of $m$ because of statistical isotropy.

In the forecasts presented in this work, we used three kinds of $\cl$: the void-void auto-correlation $\clvv$, the lensing-lensing auto-correlation $\clgg$, and the void-lensing cross-correlation $\clvg$.
We computed the $\cl$ ``tomographically''~\citep{Hu:1999ek} in a set of redshift bins, that is to say they were evaluated over two redshift bins $i$ and $j$, and were generically denoted as $\clABij$.
For example, $\clgvij$ is the spherical harmonic transform of the correlation function between the lensing signal in the $i$-th bin and the void signal in the $j$-th bin.

We evaluated $\clvv$ and $\clvg$ for $\ell \in[10,1500]$, while for $\clgg$, two scenarios were considered: a ``pessimistic'' WL scenario, with $\ell \in[10,1500]$, and an ``optimistic WL scenario, with $\ell \in[10,5000]$.

The Limber approximation~\citep{Limber1953, LoVerde2008} allowed us to write a simple integral expression for the tomographic $\cl$, where the power spectrum $\hat{P}_\textrm{AB}(k,z)$, defined in Eq.\eqref{eq:smoothedpmm}, enters the integral as
\begin{equation}
        \label{eq:cl_limber}
        \clABij \simeq \frac{c}{\Hzero} \int_{z_\text{min}}^{z_\text{max}}  \mathrm{d} z\, \frac{W_{i}^\mathrm{A}(z) W_{j}^\mathrm{B}(z)}{E(z) r^{2}(z)} \hat{P}_\textrm{AB}\left[\frac{\ell+1/2}{r(z)}, z\right] \,,
\end{equation}
where $W_{i}^\mathrm{A/B}(z)$ are suitable ``weight'' functions for the different probes $\mathrm{A}$ and $\mathrm{B}$ defined in Eqs.~\eqref{eq:w_void} and \eqref{eq:w_lensing}, $E(z)$ is the dimensionless Hubble parameter,
\begin{align}
\label{eq:hubble}
    E(z) &= \frac{\Hz}{\Hzero}\nonumber\\
    &=\sqrt{\Omm(1+z)^{3}+(1-\Omm)(1+z)^{3\left(1+w_{0}+w_{a}\right)} \mathrm{e}^{-3 w_{a} z/(1+z)}}\,,
\end{align}
 and $r(z)$ is the comoving distance,
\begin{equation}
r(z) = \frac{c}{\Hzero} \int_0^z \frac{\text{d}z'}{E(z')}\,.
\end{equation}
Throughout the forecast, the integral expression of $\clABij$ was computed numerically with the Simpson method~\citep{Brew1979}.
The signals $\clvvij$, $\clggij$, and $\clgvij$, evaluated using the reference cosmology in Table~\ref{table:par_ref}, are shown in Fig.~\ref{fig:cl_plots}.

In the \Euclid photometric survey, the minimum and the maximum redshifts are $z_\text{min}=0.001$ and $z_\text{max}=2.5$, respectively.
The $\cl$ and the distributions of galaxies and voids are sampled in ten redshift bins $i$, with the following endpoints:
\begin{align}
\label{eq:tomo_bins}
        \{ z_{e} \}=\{&0.001, \, 0.418, \, 0.560, \, 0.678, \, 0.789, \, 0.900, \, 1.019, \, 1.155,\nonumber\\
        &1.324, \, 1.576, \,2.50\}\, .
\end{align}
These bins were chosen to be equi-populated in the photometric galaxy catalog. Although, as explained in Sect.~\ref{sec:cl}, the mock void catalog was split in ten ``equi-spaced'' redshift bins, we preferred to convert the redshift void distribution to the same binning adopted for galaxies.

In order to simplify the notation of the integrand in Eq.~\eqref{eq:cl_limber}, we defined the ``kernel'' functions as
\begin{equation}
        K_{i j}^\mathrm{AB}(z)= \frac c {\Hzero} \frac{W_i^\mathrm{A}(z)W_j^\mathrm{B}(z) }{E(z) r^2(z)}\, ,
\end{equation}
with $i$, $j$ being the tomographic indices, and $\mathrm{A}$, $\mathrm{B}$ the considered probes.

The tomographic void-void angular power spectra $\clvvij$ were computed as
\begin{equation}
        \label{eq:clvv}
        \clvvij =\int_{z_\text{min}}^{z_\text{max}}\text{d}z\,K_{i j}^\mathrm{vv}(z) \hat{P}_\textrm{mm}\left[\frac{\ell+1 / 2}{r(z)}, z\right]\,,
\end{equation}
where $\hat{P}_\textrm{mm}(k,z)$ is defined in Eq.~\eqref{eq:bvsq} and we omitted the shot-noise additive term $1/\overline{n}_{i}^\mathrm{v}$ defined in Eq.~(\ref{eq:shot_noise_def}).

 The void weight function is defined as
\begin{equation}
\label{eq:w_void}
        W_{i}^{\mathrm{v}}(z) = \frac{\Hz}{c} n^\mathrm{v}_{i}(z)\,\bveffi\, .
\end{equation}

The void effective bias, $b_{\mathrm{veff}, i}$, entering in the void weight function, is given by $b_{\mathrm{veff}}(z)$, as defined in Eq.~\eqref{eq:bvz}, evaluated in the center of the $i$-th tomographic bin. The un-normalized projected void density distribution in redshift was measured directly from the \Euclid Flagship mock photometric galaxy catalog\footnote{Euclid Collaboration, in preparation.}, as explained in Sect.~\ref{sec:cl}.

The tomographic void-lensing angular power spectra, $\clgvij$, and the lensing-lensing angular power spectra, $\clggij$, were computed, respectively, as

\begin{equation}
        \label{eq:clgv}
        \clgvij =\int_{z_\text{min}}^{z_\text{max}}\text{d}z\,K_{i j}^{\mathrm{v}\gamma }(z) \hat{P}_\textrm{mm}\left[\frac{\ell+1 / 2}{r(z)}, z\right]\,
\end{equation}
and 
\begin{equation}
        \clggij  = \int_{z_\text{min}}^{z_\text{max}}\text{d}z\, K_{i j}^{\gamma \gamma}(z) \pmlimb \,,
        \label{eq:clgg}
\end{equation}
where in $\clggij\ $ we omitted the shape-noise term given by Eq.~(\ref{eq:shot_noise_def}). Following \citetalias{EuclidCollaboration2019}, in the kernels of Eqs.~(\ref{eq:clgv})-(\ref{eq:clgg}), the lensing weight function reads

\begin{equation}
        W_{i}^{\gamma}(z) = \frac{3}{2} \left( \frac{\Hzero}{c}\right) ^2 \Omm (1+z) r(z)\tilde{W}_i^\gamma(z)\,,
\label{eq:w_lensing}
\end{equation}
with
\begin{equation}
        \tilde{W}_i^\gamma(z)=\int_{z}^{z_{\max }} \mathrm{d} z^{\prime}\,n_{i}^g\left(z^{\prime}\right)\left[1-\frac{r(z)}{r\left(z^{\prime}\right)}\right]\,.
\end{equation}
Here $\tilde{W}_i^\gamma(z)$ is the lensing efficiency, defined as an integral expression of the observed galaxy distribution $\nigz$.

The density distribution of the ``observed'' galaxies in the $i$-th tomographic bin, $\nigz$, was computed as a convolution of the galaxy distribution, $\ngz$, and the photometric instrument response, as described below. According to \citetalias{EuclidCollaboration2019}, the galaxy redshift distribution, $\ngz$, adopted in this work is
\begin{equation}
        \label{eq:ngz}
        \ngz \propto \left(\frac{z}{z_{0}}\right)^{2} \exp \left[-\left(\frac{z}{z_{0}}\right)^{3 / 2}\right]\,,
\end{equation}
where $z_0={0.9}/{\sqrt{2}}$~\citep{laureijs2011euclid}. 
Redshift measurements are affected by experimental and reconstruction effects.
We modeled the probability $\pzpz$ that a true galaxy redshift $z$ will be reconstructed with a photometric redshift $z_\mathrm{p}$ as
\begin{align}
\pzpz \,& = \, \frac{1-f_{\mathrm{out}}}{\sqrt{2 \pi} \sigma_{\mathrm{b}}(1+z)} \exp \left\{-\frac{1}{2}\left[\frac{z-c_{\mathrm{b}} z_{\mathrm{p}}-z_{\mathrm{b}}}{\sigma_{\mathrm{b}}(1+z)}\right]^{2}\right\} \,  \nonumber\\
&+ \, \frac{f_{\mathrm{out}}}{\sqrt{2 \pi} \sigma_{\mathrm{o}}(1+z)} \exp \left\{-\frac{1}{2}\left[\frac{z-c_{\mathrm{o}} z_{\mathrm{p}}-z_{\mathrm{o}}}{\sigma_{\mathrm{o}}(1+z)}\right]^{2}\right\}
\label{eq:pzpz}\,.
\end{align} 

\begin{table}
        \caption{Parameters used in the photometric redshift distribution $\pzpz$ of Eq.~\eqref{eq:pzpz}}
        \centering
        \begin{tabular}{  c | c| c |c |c |c |c  }
        \toprule
                {$c_{b}$} & {$z_{b}$} & {$\sigma_{b}$} & {$c_{o}$} & {$z_{0}$} & {$\sigma_{o}$} & {$f_{\text {out }}$} \\
                \midrule
                1.0 & {0.0} & {0.05} & {1.0} & {0.1} & {0.05} & {0.1}  \\  
                \bottomrule     
        \end{tabular}
        \label{tab:pzpz}
\end{table}

\begin{figure*}[tbp]
\hspace*{-0.85cm}
\begin{tabular}{c@{}c}
             \includegraphics[width=0.52\textwidth]{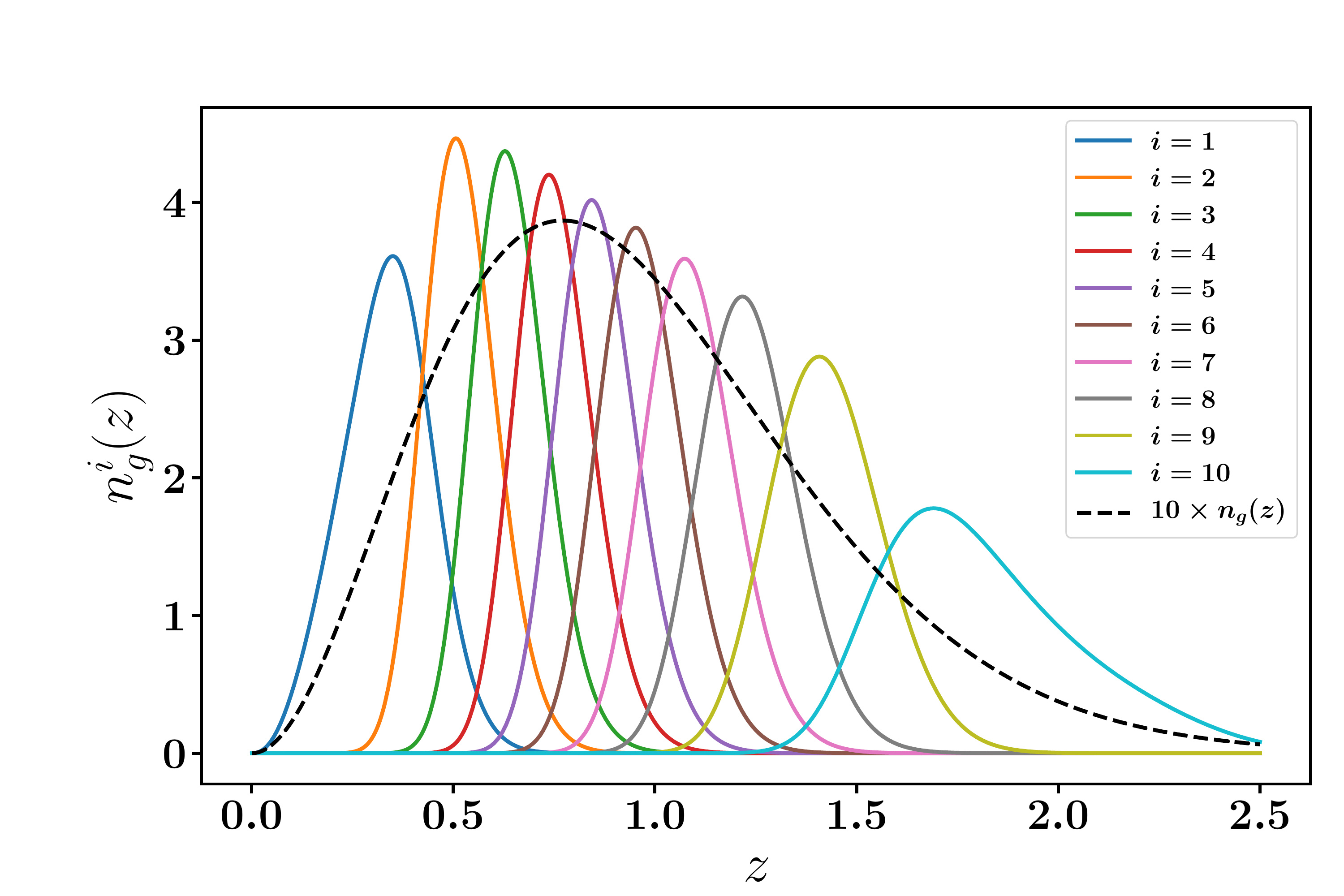}&  \includegraphics[width=0.52\textwidth]{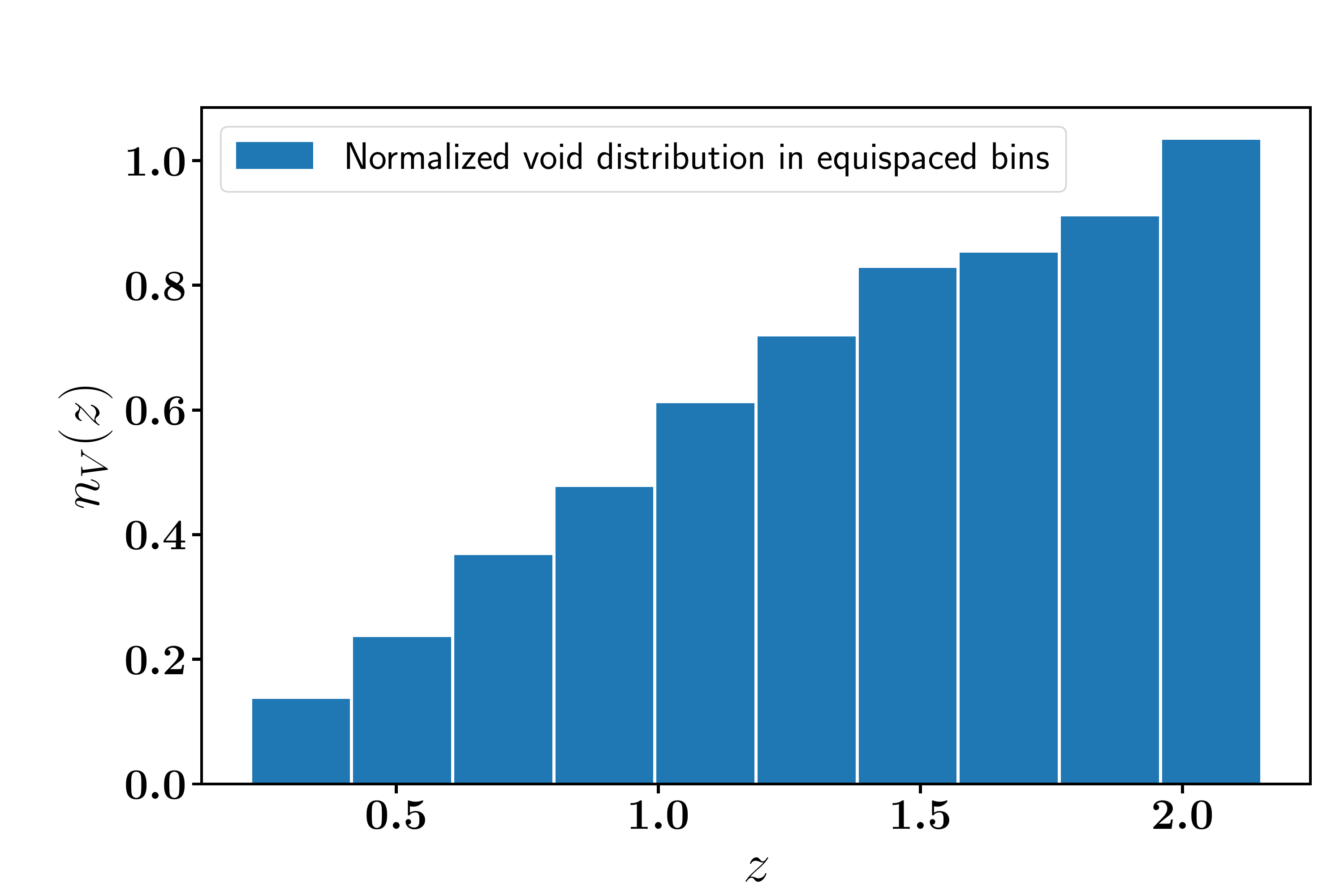}
        \end{tabular}
        \caption{Galaxy and void distributions used in this work. LEFT: Dimensionless normalized projected galaxy distributions $\nigz$ over the tomographic bins $i$. RIGHT: Dimensionless normalized projected void distribution in equi-spaced redshift bins. The corresponding un-normalized projected galaxy and void distributions are reported in Table~\ref{table:par_noise} in units of ${\rm arcmin}^{-2}$.}
        \label{fig:nigz}
        \label{fig:voids_distribution}
\end{figure*}

The modeling and the parameters entering $\pzpz$ are reported in Table~\ref{tab:pzpz}, and were taken from~\citetalias{EuclidCollaboration2019}.
The convolution of $\ngz$ and $\pzpz$ estimates the galaxy redshift distribution as measured by NISP in the photometric mode, together with ground-based observation.
The density distribution, $\nigz$, of the ``observed'' galaxies in the $i$-th tomographic bin is obtained convolving $\pzpz$ with $\ngz$ in the $i$-th bin
\begin{equation}
        \nigz = \frac{\int_{z_{i}^{-}}^{z_{i}^{+}} \mathrm{d} z_{\mathrm{p}}\, \ngz p_{\mathrm{ph}}\left(z_{\mathrm{p}} | z\right)}{\int_{z_{\mathrm{min}}}^{z_{\mathrm{max}}} \mathrm{d} z\, \int_{z_{i}^{-}}^{z_{i}^{+}} \mathrm{d} z_{\mathrm{p}} \ngz p_{\mathrm{ph}}\left(z_{\mathrm{p}} | z\right)}\,,
        \label{eq:nigz}
\end{equation}
where $z_i^+$ and $z_i^-$ are the edges of the $i$-th bin as defined in Eq.~\eqref{eq:tomo_bins}. The density distributions, $\nigz$, in the different tomographic bins are shown in the left panel of Fig.~\ref{fig:nigz}.

The matter power spectrum, $\pmkz$, was computed numerically using as Boltzmann solver the ``Code for Anisotropies in the Microwave Background''\footnote{https://camb.info/} (\camb)~\citep{Lewis1999}\footnote{The \camb\ version used in this work is \texttt{v1.0.4}.
The different \camb\ options were set to include neutrino effects and also nonlinear evolution via the Takahashi \texttt{halofit} recipe~\citep{Takahashi2012a, Bird2011}}.
In order to obtain the cosmological parameter forecasts presented in this work, we computed the matter power spectra, $\pmkz$, for each set of input cosmological parameters, which were varied with respect to the fiducial cosmology in Table~\ref{table:par_ref} to also compute the derivatives entering the Fisher matrix approach \citep{Fisher1935}.
\camb\ evaluates $\pmkz$ over a $k$-$z$ grid. According to~\citetalias{EuclidCollaboration2019}, the range of redshifts chosen in the present analysis is $z \, \in \, [0.001, \, 2.5]$, the range of wave numbers is $k\, \in \, [10^{-5}, \, 400]$ $\si{Mpc^{-1}}$, and at least 300 equispaced points in redshift are used.
\camb\ performs a logarithmic $k$-binning. In the forecast, at least 50 steps per decade are used.
The impact of different $k$ and $z$ binnings on the final results is reported in Sect.~\ref{sec:syst}.

Since both $\nigz\ $ and $\nivz$ are normalized quantities, the variation of the cosmological parameters in the Fisher forecast computation does not (or negligibly) affect these quantities. In fact, the main effect on number counts from changing the model cosmology is due to the associated volume change, which cancels out for normalized quantities. Instead, the shot-noise matrix in Eq.~(\ref{eq:shot_noise_def}) is un-normalized and computed at the fiducial cosmology assumed in the Fisher forecast computation.
The cosmological parameter variation is needed for the computations of the $\cl$ derivatives, and enters the matter power spectra, $\pmkz$, the growth factor, $G(z)$, the Hubble parameter, $\Hz$, and the comoving distance, $r(z)$, where the latter two terms change the weight functions $W_i^{\mathrm{A}}$.


\begin{table*}
\begin{adjustwidth}{-0.8cm}{}
        \caption{Survey specifications entering Eqs.~\eqref{eq:single_probe_covariance}--\eqref{eq:shot_noise_def}. Here $\bar{n}_i^g$ and $\bar{n}_i^\textrm{v}$ are in units of $\text{arcmin}^{-2}$ .}
        \centering
        \setlength{\tabcolsep}{2.pt}
        \begin{tabular}{ c | c | c | c | c | c | c | c | c | c | c | c | c }
        \toprule
        $f_\text{sky}$&$\sigma_\epsilon$&$\bar{n}_i^g$ &$\bar{n}_1^\textrm{v}\times10^{4}$ &$\bar{n}_2^\textrm{v}\times10^{4}$ &$\bar{n}_3^\textrm{v}\times10^{4}$ &$\bar{n}_4^\textrm{v}\times10^{4}$ &$\bar{n}_5^\textrm{v}\times10^{4}$ &$\bar{n}_6^\textrm{v}\times10^{4}$ &$\bar{n}_7^\textrm{v}\times10^{4}$ &$\bar{n}_8^\textrm{v}\times10^{3}$ &$\bar{n}_9^\textrm{v}\times10^{3}$ &$\bar{n}_{10}^\textrm{v}\times10^{3}$ \\
        \midrule
        0.3636 & 0.3 & $3$& $3.03$& $3.52$& $3.68$& $4.09$& $4.7$& $5.71$& $7.39$& $1.05$ & $1.82$& $9.36$\\
        \bottomrule     
        \end{tabular}
        \label{table:par_noise}
        \end{adjustwidth}
\end{table*}

In order to use realistic estimates of void bias and void distribution from the \Euclid photometric galaxy sample, in the present analysis we employed the Flagship mock galaxy catalog~(\Euclid Collaboration, in prep). This is based on an $N$-body simulation of $12\,600^3$ DM particles in a periodic box of 3780 $\hMpc$ on a side~\citep{Potter2017}, with a flat $\lcdm$ cosmology very similar to that reported in Table~\ref{table:par_ref}, namely $\Omm = 0.319, \Omb = 0.049, \OmL = 0.681, \sige = 0.83,$ and $\ns = 0.96, h = 0.67$, as obtained by \textit{Planck}~\citepalias{Collaboration2016}.
Dark matter halos were identified with the ROCKSTAR halo finder~\citep{Behroozi2013}, and populated with central and satellite galaxies using a halo occupation distribution (HOD) framework to reproduce the relevant observables for the \Euclid main cosmological probes. The HOD algorithm~\citep{Carretero2015, Crocce2015} was calibrated exploiting several observational constraints, such as the local luminosity function for the faintest galaxies~\citep{Blanton2003, Blanton2005} and GC statistics as a function of luminosity and color~\citep{Zehavi2011}. The resulting Flagship galaxy mock lightcone spanned one octant of the sky and simulated both spectroscopic and photometric \Euclid galaxy samples. In this paper we consider the latter, which extends up to redshift $z=2.3$ and in which a Gaussian photometric redshift error of $\Delta z=0.05(1+z)$ was applied to each galaxy.

To identify cosmic voids in this catalog, we applied the 2D void finder of \cite{Sanchez2017, Vielzeuf2019} to the photometric galaxy sample, computing the number of voids in ten ``equi-spaced'' redshift bins, as shown in the right panel of Fig.~\ref{fig:voids_distribution}, where the ratio between the void number in each bin and the bin width represents the void projected density at the bin center. We find that the 2D void population traced by photometric galaxies in the Flagship catalog extends from $r_{\rm{ v, min}} \sim 25\,\hMpc$ up to $r_{\rm{v, max}} \sim 300\,\hMpc$, and its projected spatial density, $n^\textrm{v}(z)$, is obtained by interpolating the void density in each bin center.

Moreover, to verify that the bias modeling, Eq.~(\ref{eq:bvz}), used in our forecasts was representative of \Euclid observations, we measured the void bias redshift evolution in the obtained Flagship photometric void catalog.  
To this aim, we followed the methodology presented in \cite{Hamaus2014a}. We used the open-source code \texttt{nbodykit}~\citep{hand2018} to compute the void auto power-spectrum in eight redshift bins ($z=[0.3,0.5,0.75,1.0,1.25,1.5,1.75,2.]$) with a bin size of $\Delta z=0.2$. Then, from Eq.~(\ref{eq:bvsq}), we inferred the void bias as
\begin{equation}
\bveff(z)=\sqrt{\frac{P_\textrm{vv}(k,z)-1/\bar{n}^{\textrm{v}}}{P_{\textrm{mm}}(k,z)}}\,,
\label{eq:bias_meas}
\end{equation}
where the void power spectrum was measured in the Flagship photometric void catalog.
Then, we averaged the measured bias in the range $0.05<k({\hMpc})<0.1$, that it to say, in the regime in which the void bias is constant as a function of the scale~ \citep[see][]{Hamaus2014a}.

\begin{figure}[tbp]
    \centering
    \includegraphics[width=1\columnwidth]{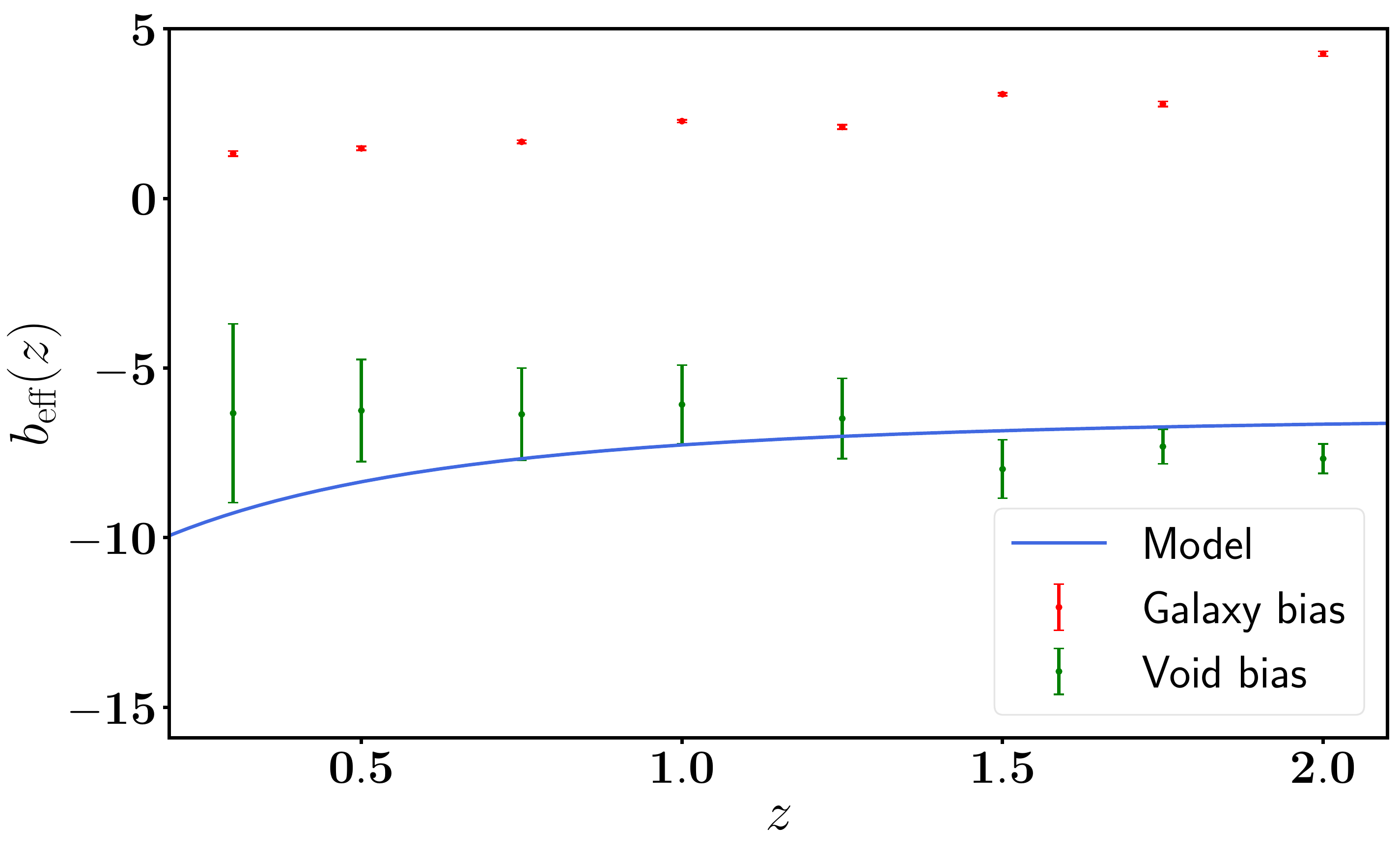}
    \caption{Effective void bias (green error bars) and galaxy bias (red error bars) as measured in the Flagship simulation versus the theoretical effective void bias (blue line) of Eq.~\eqref{eq:bvz} .
    Void bias measurements have a bigger error bar, compared to galaxy bias measurements, since there are fewer voids than galaxies.}
    \label{fig:void_bias}
\end{figure}

The evolution of the void bias as a function of redshift in the Flagship photometric void catalog (green error bars) is shown in Fig.~\ref{fig:void_bias}.
The measurement errors were estimated using a jackknife resampling. The red stars in Fig.~\ref{fig:void_bias} show the galaxy bias measured from the Flagship mock catalog using Eq.~\eqref{eq:bias_meas}. 
Finally the solid blue line shows the theoretical bias modeling, Eq.~\eqref{eq:bvz}, obtained via the PBS formalism and adopted for our forecast computation. We note that the measured negative void bias shows a redshift evolution close enough to the theoretical one. Therefore, we consider the PBS prescription of the void bias evolution acceptable for the purposes of the forecast analysis presented in this work. We accounted for any theoretical uncertainty on the void threshold and size function by marginalizing over the normalization of the effective bias in the pessimistic case. 
We verified that the agreement between the theoretical void size function and the one measured from the photometric void sample in the Flagship simulation was good enough for the forecasting purposes of this work, and we postpone to a future work a detailed calibration against simulated data, when more realistic mocks, accounting for detailed Euclid survey specification and systematics, will be available.


\section{The void-lensing Fisher matrix}
\label{sec:fisher}

The Fisher matrix formalism \citep{Fisher1935} is used to predict uncertainties of cosmological parameter measurements from different probes.
This section summarizes the approach and provides details on its application to the cases considered in this work.

The Fisher matrix is defined as the expectation value of the second derivatives of the logarithm of the likelihood $L$:
\begin{equation}
\label{eq:fisher}
\fisher = - \left \langle \frac{\partial^2 \ln L}{\partial \alpha \partial \beta} \right \rangle,
\end{equation}
where $\alpha$ and $\beta$ are the parameters of interest.
The expected ``error covariance matrix'' is the inverse of the Fisher matrix:
\begin{equation}
\mathcal{C}_{\alpha \beta} = (F^{-1})_{\alpha \beta}\, .
\end{equation}
The diagonal elements of the error covariance matrix  are the squares of the marginalized 1-$\sigma$ errors on the parameters:
\begin{equation}
    \sigma_{\alpha} = \sqrt{\mathcal{C}_{\alpha\alpha}}\,.
\end{equation}
As we deal with angular power spectra, it is convenient to introduce the matrix $\Sigma^{\mathrm{AB}}_{ij}$ associated with a given $\cl:$
\begin{equation}
\label{eq:single_probe_covariance}
    \Sigma^{\mathrm{AB}}_{ij}(\ell) = 
\sqrt{\frac{2}{(2\ell+1)\Delta\ell \,f_{\rm sky}}}
\left[ 
C_{ij}^{AB}(\ell) + N_{ij}^{AB}(\ell)
\right]\,,
\end{equation}
where $\Delta\ell$ is the multipole bin width, and $f_ \mathrm{sky}$ the sky fraction covered by the survey.  The ``shot-noise matrix'', $N^{\mathrm{AB}}_{ij}(\ell)$, depends on the particular probe combination. For the $\cl$ of the void clustering, WL , and void-lensing, assuming Poisson statistics, they are respectively
\begin{equation}
        N_{ij}^\textrm{vv}=\frac{1}{\bar{n}^\textrm{v}_i}\delta_{ij}\, ,
        \qquad
        N_{ij}^{\gamma\gamma}=\frac{\sigma_\epsilon^2}{\bar{n}^g_i}\delta_{ij}\, ,
        \qquad
        N_{ij}^{\gamma\textrm{v}}=N_{ij}^{\textrm{v}\gamma}=0\, ,
        \label{eq:shot_noise_def}
\end{equation}
where $\delta_{ij}$ is the Kronecker delta,  $\sigma_\epsilon^2$ is the galaxy shape noise, and $\bar{n}_i^g$ and $\bar{n}_i^\textrm{v}$ are the un-normalized average galaxy and void surface densities in the $i$-th tomographic bin computed in the fiducial cosmology, respectively.
The survey specifications used to compute the covariances are shown in Table~\ref{table:par_noise}.

In the case of a single probe ($\mathrm{A}=\mathrm{B}$), the covariance matrix of the $a_{\ell m}$ is simply given by Eq.~\eqref{eq:single_probe_covariance}. When two or more probes are combined together, one needs to construct the full covariance matrix, $\mathbf{\Sigma}^{\mathrm{XC}}$, composed of matrix blocks defined in Eq.~\eqref{eq:single_probe_covariance}.
For the probes $\gamma$ and $\mathrm{v}$ considered in this work, $\mathbf{\Sigma}^{\mathrm{XC}}$ is given by
\begin{equation}
\mathbf{\Sigma}^{\mathrm{XC}}(\ell) = 
\begin{pmatrix}
\mathbf{\Sigma}^{\gamma\gamma}(\ell) & \mathbf{\Sigma}^{\gamma \textrm{v}}(\ell) \\ 
\mathbf{\Sigma}^{\textrm{v}\gamma}(\ell) & \mathbf{\Sigma}^\textrm{vv}(\ell) 
\end{pmatrix}\,.
\end{equation}
The covariance and the $\cl$ matrix share the same structure, and therefore the block matrix reads
\begin{equation}
\mathbf{C}^{\mathrm{XC}}(\ell) = 
\begin{pmatrix}
\mathbf{C}^{\gamma\gamma}(\ell) & \mathbf{C}^{\gamma\textrm{v}}(\ell) \\ 
\mathbf{C}^{\textrm{v}\gamma}(\ell) & \mathbf{C}^\textrm{vv}(\ell) 
\end{pmatrix}\,.
\end{equation}

Assuming that the $a_{\ell m}$, defined in Eq.~\eqref{eq:alm_def}, follow a multivariate Gaussian distribution, an analytical expression for the Fisher matrix elements is given by\footnote{In Appendix \ref{sec:appendix_C} we show that this expression is valid both in the so-called ``field'' and ``estimator'' perspectives.}

\begin{equation}\label{eq:fisher_analytical}
F_{\alpha\beta} = 
\sum_{\ell=\ell_{\rm min}}^{\ell_{\rm max}} 
\operatorname{Tr}\left\lbrace
[\mathbf{\Sigma}(\ell)]^{-1} \,
\frac{\partial \mathbf{C}(\ell)}{\partial \alpha}\,
[\mathbf{\Sigma}(\ell)]^{-1}
\frac{\partial \mathbf{C}(\ell)}{\partial \beta}
\right \rbrace\,. 
\end{equation} 

The formula above applies both to the single- and two-probe correlation case, provided that the $\cl$ and the covariance matrices are chosen accordingly.
The Fisher matrix computation involves the derivatives of the $\cl$ with respect to cosmological and nuisance parameters.
In order to ensure reliable results, numerical derivatives were computed with \footnote{In an early stage of the analysis, numeric derivatives were also performed with a five-point stencil method. However, this method has shown numeric instabilities, and was therefore discarded.} a numerical approach based on the SteM fitting procedure~\mbox{
\citep{Camera2017}}\hspace{0pt}
, which is based on a iterative linear regression, and a semi-analytical approach, in which both analytical and numerical derivatives are used.

\begin{figure*}[tbp]
\centering
 \hspace*{-0.83cm}
        \begin{tabular}{l l}
             \includegraphics[width=0.73\textwidth]{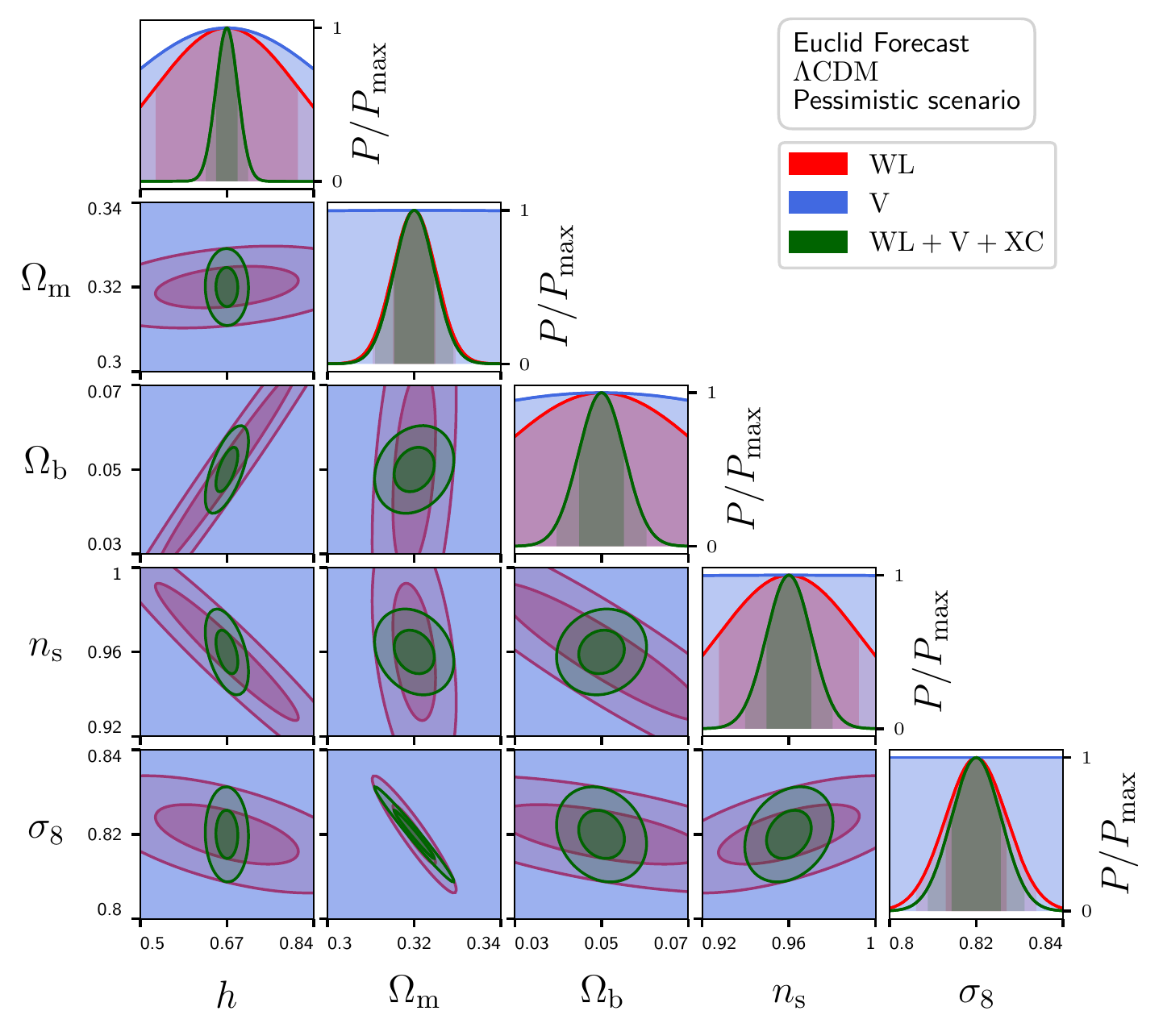}\\  \includegraphics[width=0.73\textwidth]{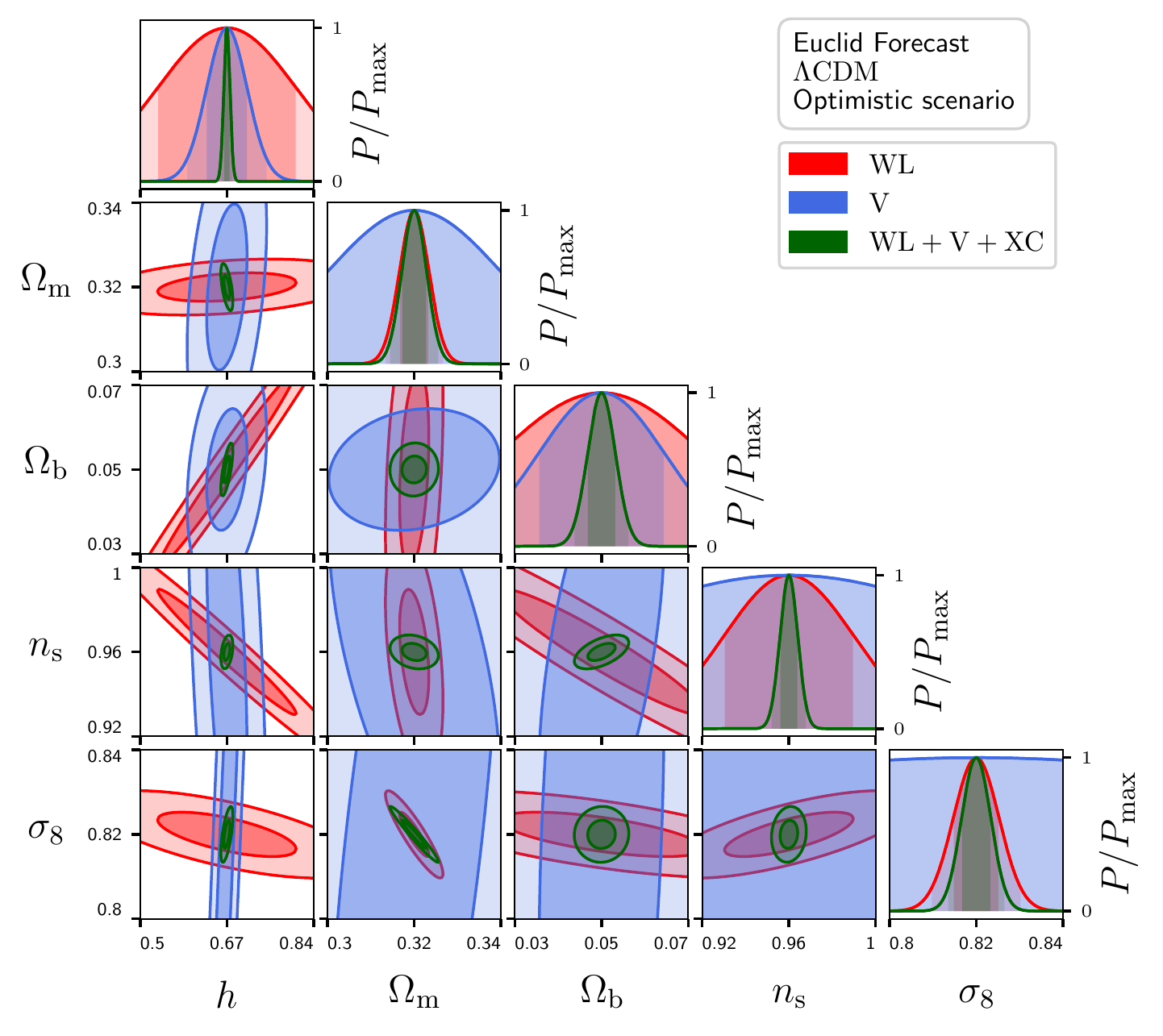}
        \end{tabular}
        \caption{Fisher matrix marginalized contours for the (baseline) $\Lambda$CDM model, in the pessimistic (\textit{top}) and in the optimistic (\textit{bottom}) scenarios.}
        \label{fig:lcdm_pess_opt}
        \end{figure*}


\begin{figure*}[tbp]
\centering
\hspace*{-0.83cm}
        \begin{tabular}{l l}
             \includegraphics[width=0.73\textwidth]{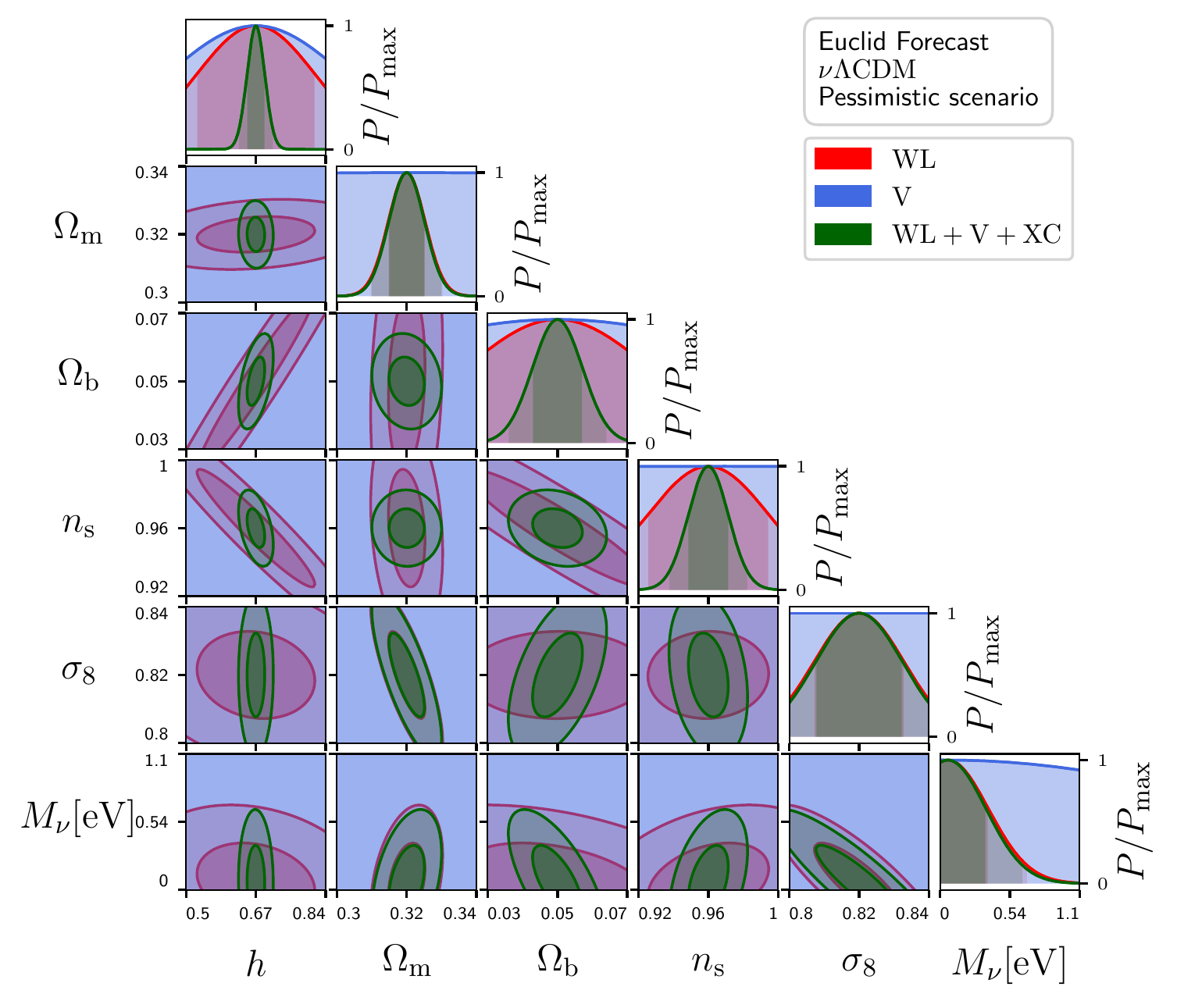}\\ \includegraphics[width=0.73\textwidth]{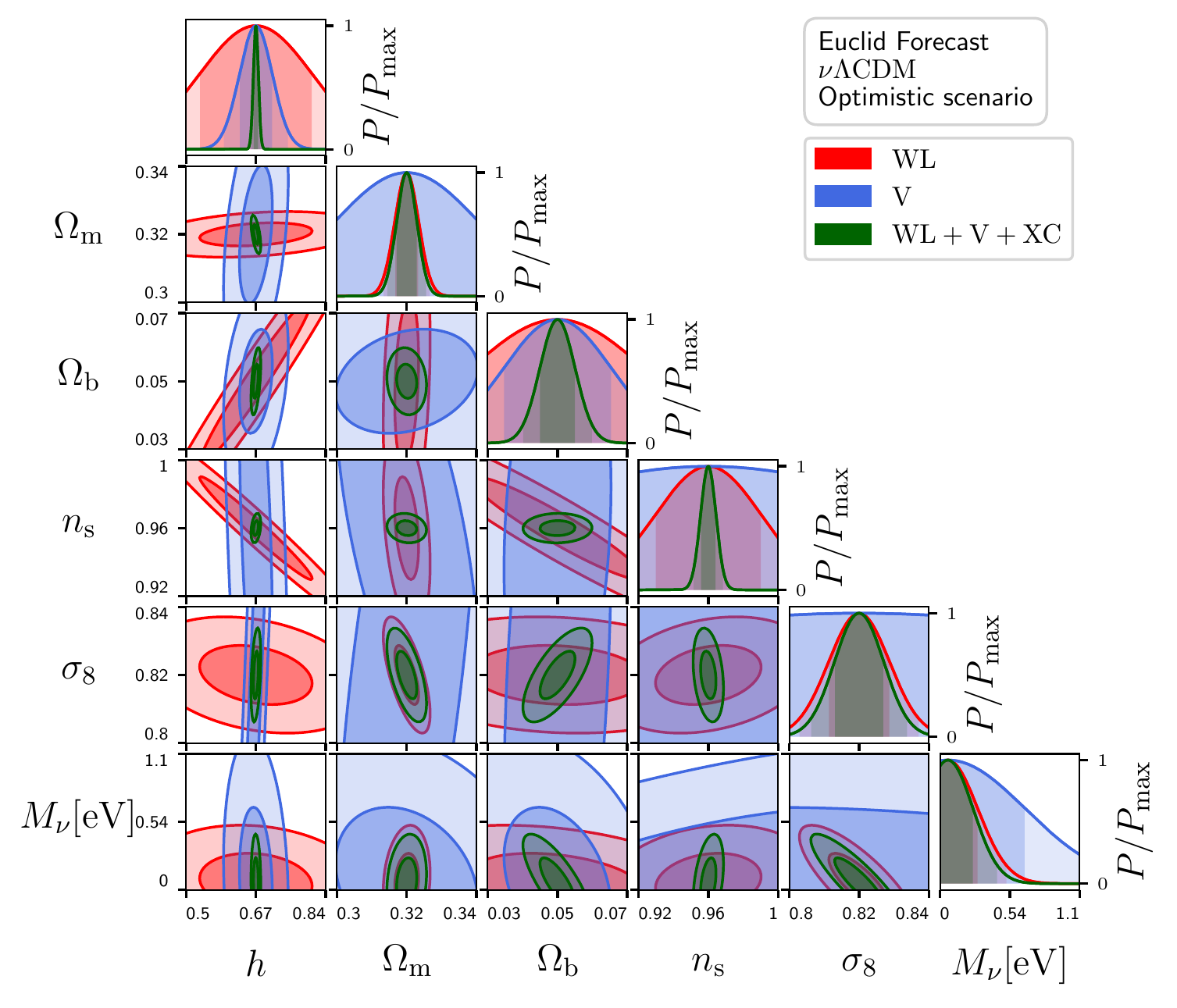}
        \end{tabular}
        \caption{Fisher matrix marginalized contours for the $\nu\Lambda$CDM model, in the pessimistic (\textit{top}) and in the optimistic (\textit{bottom}) scenarios.}
        \label{fig:nlcdm_pess_opt}
\end{figure*}

\section{Results: Correlations, errors, and the inclusion of the void bias evolution}
\label{sec:results}
\begin{figure*}[tbp]
\centering
\hspace*{-0.83cm}
        \begin{tabular}{l l}
             \includegraphics[width=0.73\textwidth]{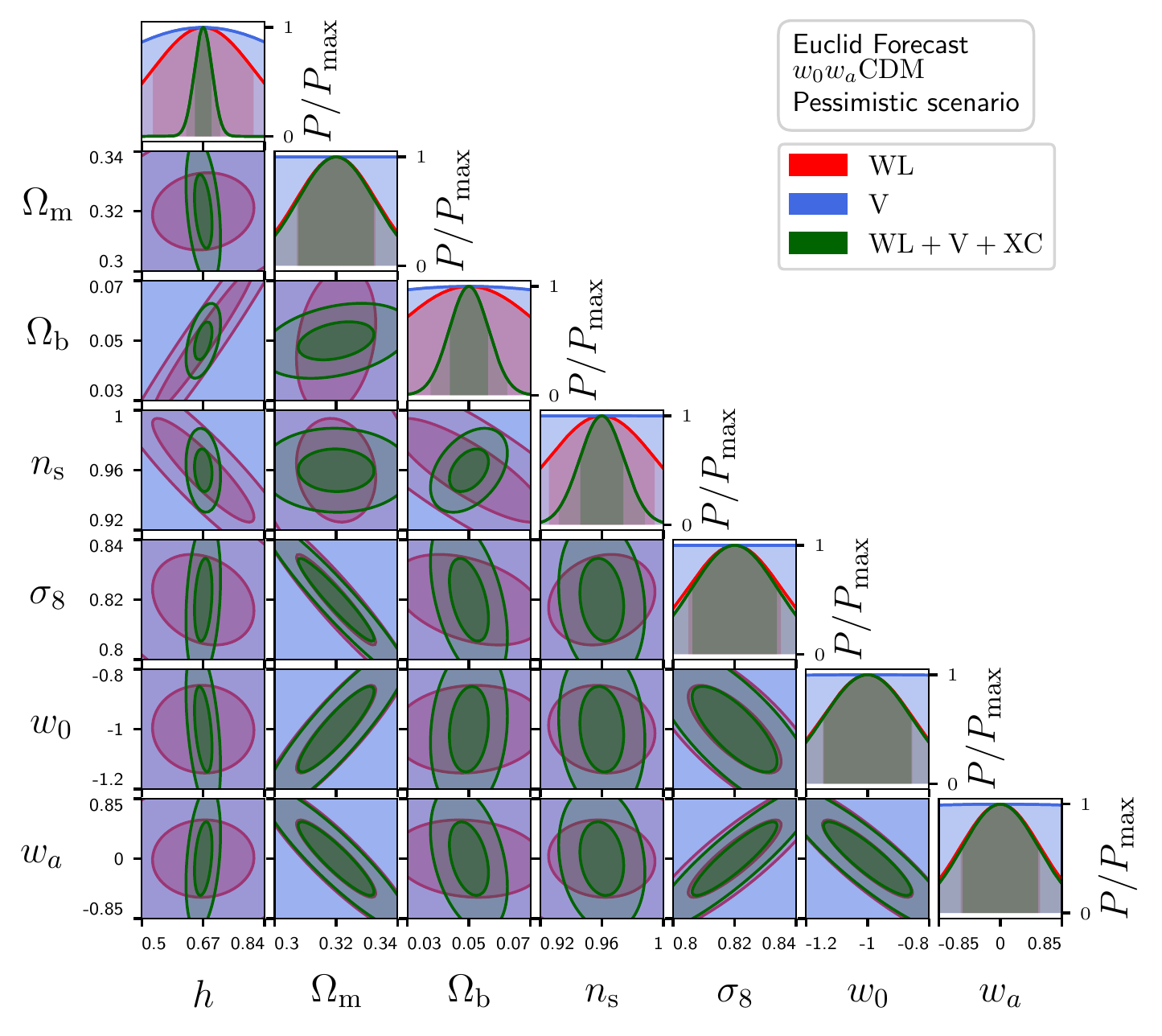}\\  \includegraphics[width=0.73\textwidth]{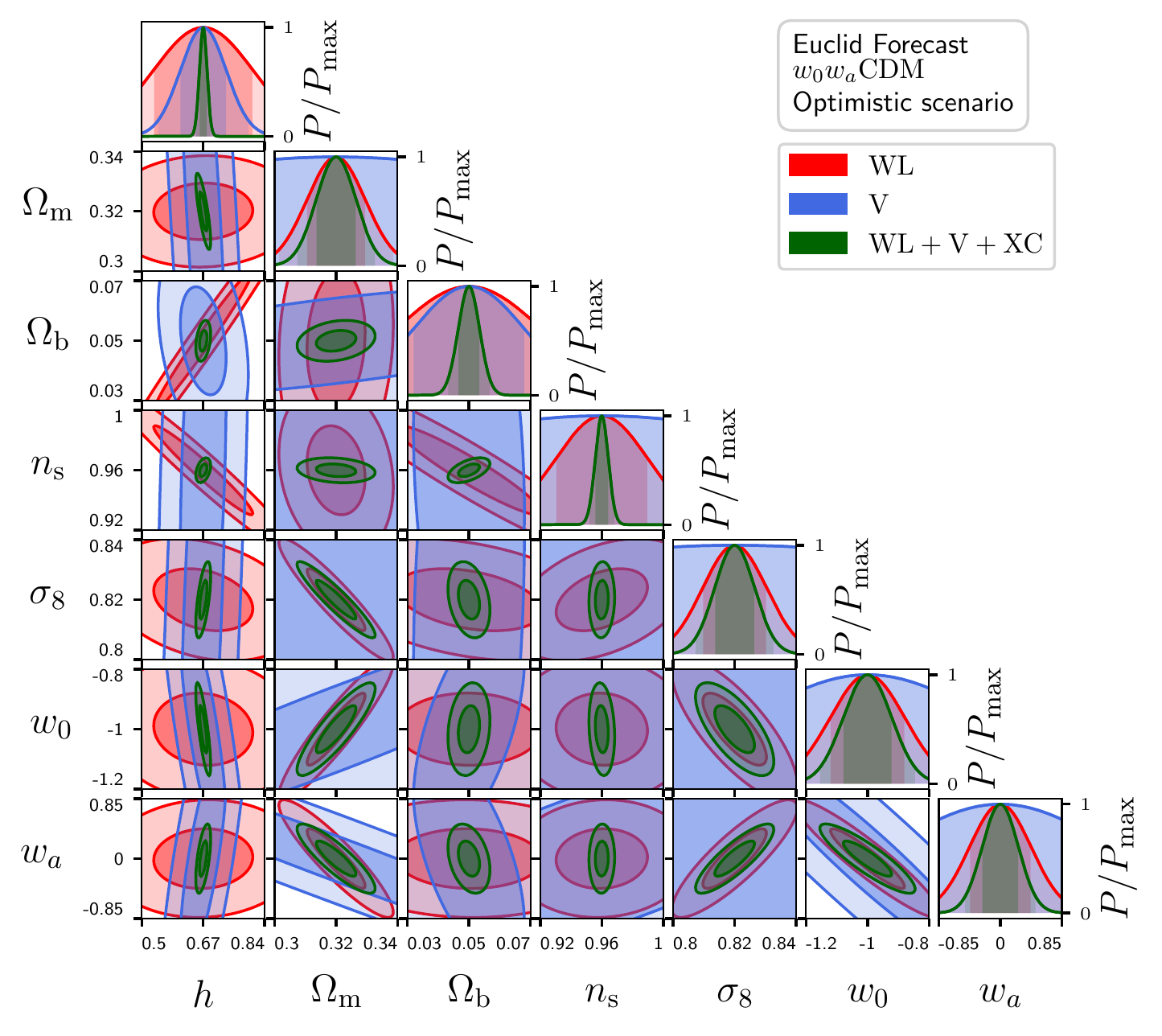}
        \end{tabular}
        \caption{Fisher matrix marginalized contours for the $\wcdm$ model, in the pessimistic (\textit{top}) and in the optimistic (\textit{bottom}) scenarios.}
        \label{fig:wcdm_pess_opt}
        \end{figure*}

In this section we present the results obtained implementing the analysis described in previous sections. We discuss the forecasts of cosmological parameters, expressed as Fisher matrix marginalized contours, obtained combining the expected \Euclid WL and angular void clustering.

We report parameter forecasts in two different scenarios, namely the pessimistic and optimistic setups, described in Sect.~\ref{sec:void} for the void bias, and in Sect.~\ref{sec:cl} for the WL, respectively. They are also summarized in Table~\ref{tab:configurations}.
The 1-$\sigma$ errors, together with the DE figure of merit\footnote{The FoM is defined as the inverse of the square root of the determinant of the covariance matrix $C_{\wz \wa}$ and is inversely proportional to the area of the contour ellipse $\wz$-$\wa$ in the marginalized parameter plane~\citepalias{EuclidCollaboration2019}. The covariance matrix is the inverse of the Fisher matrix.} (FoM), in both scenarios, and different cosmological models, are reported in Table \ref{tab:FoM_marginalised_errors_all}. The corresponding contour plots are shown in Figs.~\ref{fig:lcdm_pess_opt}--\ref{fig:nwcdm_pess_opt}. Focusing on the DE equation of state and on the total neutrino mass, the contour plots for $\wz, \wa, \Mnu$, evaluated respectively for the pessimistic and optimistic scenarios, are also shown in the top panels of Fig.~\ref{fig:nwcdm_interm_both_r}.

\subsection{The standard $\lcdm$ cosmology}
The contour plots for the $\Lambda$CDM cosmology are reported in Fig.~\ref{fig:lcdm_pess_opt}, the pessimistic case in the top panel and the optimistic case in the bottom panel. In both the pessimistic and optimistic scenarios, we can observe that WL has a larger constraining power than photometric void clustering, except for $h$ and $\Omb$ in the optimistic case, where for $h$ void clustering provides better constraints than WL by a factor of $\sim 3$, while for $\Omb$ the constraints from the two single probes look comparable. This is due to the form of the WL kernel and the integration along the line of sight, and to the presence of baryon acoustic oscillations (BAO) in the void angular auto power spectrum, $\clvv$, and the void-lensing cross spectrum, $\clvg$, (see  Fig.~\ref{fig:cl_plots}), which are instead completely washed out in the WL angular power spectrum, $\clgg$.
We also remark that, in general, both in the pessimistic and optimistic scenarios, the parameters for which the constraints improve most, when combining void clustering and WL, are not only $h$ and $\Omb$, but also $\ns$; similar conclusions (without WL) were found by~\mbox{
\cite{Kreisch2021}}.

In the case of $\ns$, this happens since the WL and void clustering ellipses in the $\ns-\Omb$ plane happen to be orthogonal. We verified that, even if $\ns$ and $\Omb$ produce an increment in the same direction for $\clgg$ and $\clvv$, which would imply that the two parameters are negatively correlated, this orthogonality comes from projecting the large parameter space onto the $\ns-\Omb$ 2D space; in other words, it is due to the Fisher matrix inversion.

Finally, when moving from the pessimistic to the optimistic scenario, all marginalized errors decrease: in particular the error on $\ns$ decreases by a factor of $\sim2$, the constraints on $\sige$ and $\Omm$ get $\sim40\%$ tighter, and the uncertainty on $h$ decreases by a factor of $\sim3$. The increase in the constraining power on $\sige$ is somehow expected: the void bias affects the overall amplitude of the $\cl$, so it is degenerate with $\sige$. On the one hand, in the pessimistic scenario, both the value of the bias parameters and $\sige$ were assumed to be measured from data, and this increases the uncertainties on $\sige$. On the other hand, in the optimistic scenario, we assumed the cosmological dependence of the effective void bias to be known and we exploited it in the forecasts of the cosmological parameters.

\begin{table*}[htbp]
\caption{Marginalized 1-$\sigma$ errors in different cosmological scenarios for the galaxy weak lensing (WL) and angular void clustering (V) probes, together with their combinations in the case they are assumed to be independent, WL+V, and when their cross-correlation is included. In each probe block, the first row shows the errors for the pessimistic scenario, while the second row corresponds to the optimistic one.}
\centering
\setlength{\tabcolsep}{2pt}
\begin{tabular}{l |c |c |c |c |c |c |c |c | c}
\toprule
Probe  
& $h$ & $\Omm$ & $\Omb$ & $\sige$ & $\ns$ & $\Mnu[\mathrm{eV}]$ & $\wz$ & $\wa$ & FoM\\
\midrule
\multicolumn{10}{c}{{$\Lambda$CDM}}\\
\midrule
\multirow{2}{*}{$\wklens$} 
& 0.141 &    0.00494 &    0.0244 &    0.00708 &  0.0327   & -- & -- & -- & --\\
& 0.136 &    0.00339 &    0.0236 &    0.00528 &  0.0298  & -- & -- & -- & --\\
\hline
\multirow{2}{*}{$\Voids$}
& 0.215 &    0.252 &    0.0625 &    0.735 &  0.581  & -- & -- & -- & --\\
& 0.0398 &    0.0198 &    0.0145 &    0.110 &  0.103  & -- & -- & -- & --\\
\hline
\multirow{2}{*}{$\wlpv$}
& 0.107 & 0.00483 & 0.0187 & 0.00655 &  0.0257 & -- & -- & -- & --\\
& 0.00735 & 0.00304 & 0.00389 & 0.00363 &  0.00432  & -- & -- & -- & --\\
\hline
\multirow{2}{*}{$\xc$}
& 0.0216 &    0.00466 &    0.00530 &    0.00575 &  0.0104  & -- & -- & -- & --\\
& 0.00604 &    0.00285 &    0.00321 &    0.00338 &  0.00409  & -- & -- & -- & --\\
\midrule
\multicolumn{10}{c}{{$\nlcdm$}}\\
\midrule
\multirow{2}{*}{$\wklens$}
& 0.144 &    0.00526 &    0.0265 &    0.0129 &  0.0348  &              0.310 & -- & -- & --\\
& 0.138 &    0.00342 &    0.0248 &    0.00873 &  0.0304  &              0.229 & -- & -- & --\\
\hline
\multirow{2}{*}{$\Voids$}
& 0.216 &    0.254 &    0.0667 &    0.812 &  0.707  &              2.46 & -- & -- & --\\
& 0.0402 &    0.0205 &    0.0154 &    0.113 &  0.131  &              0.593 & -- & -- & --\\
\hline
\multirow{2}{*}{$\wlpv$}
& 0.109 & 0.00519 & 0.0203 & 0.0127 &  0.0272  & 0.302 & -- & -- & --\\
& 0.00741 & 0.00309 & 0.00579 & 0.00752 &  0.00461  & 0.202 & -- & -- & --\\
\hline
\multirow{2}{*}{$\xc$}
& 0.0216 &    0.00509 &    0.00718 &    0.0124 &  0.0118  &              0.292 & -- & -- & --\\
& 0.00605 &    0.00289 &    0.00505 &    0.00705 &  0.00443  &              0.193 & -- & -- & --\\
\midrule
\multicolumn{10}{c}{{$\wcdm$}}\\
\midrule
\multirow{2}{*}{$\wklens$}
& 0.141 &    0.0130 &    0.0246 &    0.0152 &  0.0348  & -- &  0.147 &  0.552 &  26.9\\
& 0.138 &    0.00949 &    0.0237 &    0.0104 &  0.0299  & -- &  0.121 &  0.426 &   53.9\\
\hline
\multirow{2}{*}{$\Voids$}
& 0.316 &    0.394 &    0.0793 &    0.804 &  0.674  & -- &  2.30 &  6.59 &   0.253\\
& 0.0636 &    0.0934 &    0.0181 &    0.113 &  0.166  & -- &  0.390 &  1.55&    5.11\\
\hline
\multirow{2}{*}{$\wlpv$}
& 0.108 & 0.0128 & 0.0191 & 0.0148 &  0.0279  & -- & 0.146 & 0.543 &  27.7\\
& 0.0131 &    0.00744 & 0.00407 & 0.00748 &  0.00461  & -- &  0.0912 &  0.296 & 84.7 \\
\hline
\multirow{2}{*}{$\xc$}
& 0.0244 &    0.0124 &    0.00638 &    0.0139 &  0.0143  & -- &  0.144 &  0.526 &  29.3\\
& 0.0105 &    0.00649 &    0.00348 &    0.00650 &  0.00425  & -- &  0.0791 &  0.251 &  106\\
\midrule
\multicolumn{10}{c}{{$\nwcdm$}}\\
\midrule
\multirow{2}{*}{$\wklens$}
& 0.148 &    0.0148 &    0.0270 &    0.0273 &  0.0430  &              0.413 &  0.149 &  0.630 &  20.2\\
& 0.139 &    0.00950 &    0.0250 &    0.0120 &  0.0305  &              0.231 &  0.122 &  0.429 &   53.6\\
\hline
\multirow{2}{*}{$\Voids$}
& 0.351 &    0.482 &    0.0801 &    1.26 &  0.710  &              4.21 &  2.43 &  7.98 &   0.148\\
& 0.0676 &    0.0987 &    0.0182 &    0.119 &  0.169  &              0.655 &  0.409 &  1.67 &    4.63\\
\hline
\multirow{2}{*}{$\wlpv$}
& 0.110  &      0.0146 & 0.0204 & 0.0267 &  0.0341  & 0.397 & 0.148     & 0.614 &  20.9\\
& 0.0131& 0.00747 & 0.00584  &  0.0102 &  0.00492  &  0.203 &  0.0912 & 0.296 & 84.4  \\
\hline
\multirow{2}{*}{$\xc$}
& 0.0244 &    0.0142 &    0.00721 &    0.0259 &  0.0200  &              0.380 &  0.145 &  0.591 &  22.6\\
& 0.0105 &    0.00652 &    0.00514 &    0.00929 &  0.00463  & 0.182 &  0.0791 &  0.251 &  105\\
\bottomrule
\end{tabular}
\label{tab:FoM_marginalised_errors_all}
\end{table*}
\subsection{The massive neutrino cosmology}
Here we present the results in the case where the total neutrino mass, $\Mnu$, is not kept fixed to its fiducial value, but it is a quantity whose value has to be determined from the data. Strictly speaking, in the case of the neutrino mass, the Fisher matrix approach goes beyond its range of validity, as the associated likelihood is non-Gaussian being truncated at $\Mnu=0$. However, as shown in Fig. 1 of~\mbox{
\cite{Brinckmann:2018cvx}}\hspace{0pt}
, the posterior obtained from the MCMC is in good agreement with the one obtained from the truncated Fisher matrix.
Furthermore, we notice that, in the case of $\wklens$, the neutrino constraints presented in this work do not represent the final $\Euclid$-$\wklens$ constraints, as intrinsic alignments are not taken into account in this analysis. Moreover, a varying $\Mnu$ is not considered when the spectroscopic galaxy clustering, $\text{GC}_\text{sp}$, is included in Sect.~\ref{GCs_r25}. The contour plots for this cosmology are reported in Fig.~\ref{fig:nlcdm_pess_opt}, with the pessimistic case on the left and the optimistic one on the right. On the one hand, adding $\Mnu$ as a free parameter mainly impacts the measurement of $\sige$, weakening its constraints by a factor of $\sim 2$ with respect to the baseline $\Lambda$CDM case. This is expected since the effects of $\sige$ and massive neutrinos on the matter power spectrum are similar: the former regulates the normalization of $P_{\textrm{mm}}$, and the latter suppresses $P_{\textrm{mm}}$ in a scale-dependent way, due to neutrinos free streaming~\citep{Lesgourgues2006}.
On the other hand, such a suppression allows void clustering to impact positively on the neutrino mass measurements: the error on $\Mnu$ decreases with respect to the WL case alone, by $\sim5\%$ and $\sim15\%$ in the pessimistic and optimistic scenarios, respectively.
In addition to considering a fiducial total mass of $\Mnu=\SI{0.06}{eV}$ in the normal hierarchy scenario, we also computed parameter forecasts assuming a neutrino mass degenerate scenario with fiducial total mass $\Mnu=\SI{0.15}{eV}$\footnote{The value $\Mnu=\SI{0.15}{eV}$ is compatible with either an inverted or a normal hierarchy~\citep{Jimenez_etal_2010}; we chose a normal hierarchy in this case, as implemented in \citetalias{EuclidCollaboration2019}.}, from the best-fit value in \cite{Pellejero-Ibanez2016}. With this choice, the marginalized errors on $\Mnu$ decrease by $\sim 5\%$. This is expected: the higher the neutrino mass, the more the matter power spectrum is sensitive to its effects~\citep{Lesgourgues2006}.

\subsection{The dynamical dark energy cosmology}
In this cosmological scenario, we analyzed the $\wcdm$ model, considering a time-dependent DE equation of state. The contour plots for this cosmology are reported in Fig.~\ref{fig:wcdm_pess_opt}, both in the pessimistic and optimistic scenarios. Again, adding $\wz$ and $\wa$ mainly affects $h$, $\sige$, and $\Omm$, increasing their errors. The impact on $h$ and $\Omm$ is explained considering that $\wz$ and $\wa$ enter the Hubble parameter $H(z)$ \eqref{eq:hubble}, where the dependence from the DE equation of state has been exploited. Moreover, $\wz$ and $\wa$ also enter the linear growth factor~\citep{Linder2003}, and therefore impact  $\Omm$ and the normalization $\sige$ again.
When moving from the pessimistic to the optimistic scenario, the constraints on $\wz$ and $\wa$ improve, and consequently the FoM is enhanced by a factor of $\sim$ 3. This can be explained since, in the optimistic scenario, we exploited both the cosmology dependence of the growth factor in the void bias evolution, $b_{\rm v}$, as well as the cosmology dependence of the void size function used to compute the effective void bias in Eq.~\eqref{eq:bvz}. This is confirmed by previous works~\citep{Pisani2015, Verza2019b}, although these focused on upcoming spectroscopic data and found that the void size function is sensitive to the DE equation of state.
We also stress the increase in the constraining power when the two probes are combined together: comparing from Table~\ref{tab:FoM_marginalised_errors_all} the constraints produced by WL alone against the ones obtained from its combination with the void clustering and void-lensing cross-correlation, we can observe that the FoM is enhanced by $\sim10\%$ in the pessimistic scenario, and by a factor of $\sim2$  in the optimistic one.

Finally, in order to measure the impact on parameter forecasts of the void-lensing cross-correlation signal, we also evaluated the marginalized errors when WL and void clustering are treated as independent probes, that is, when the Fisher matrices of the single probes are directly summed up. Then, we compared the FoM against the case when the void-lensing cross-correlation is included in the analysis. We found that, in the latter case, the DE FoM is enhanced by $5\%$ in the pessimistic scenario, and by $20\%$ in the optimistic one, as reported in Table~\ref{tab:FoM_marginalised_errors_all}.

\subsection{Including the spectroscopic galaxy clustering}
\label{GCs_r25}

\begin{figure*}[tbp]
\centering
\hspace*{-0.83cm}
        \begin{tabular}{l l}
             \includegraphics[width=0.73\textwidth]{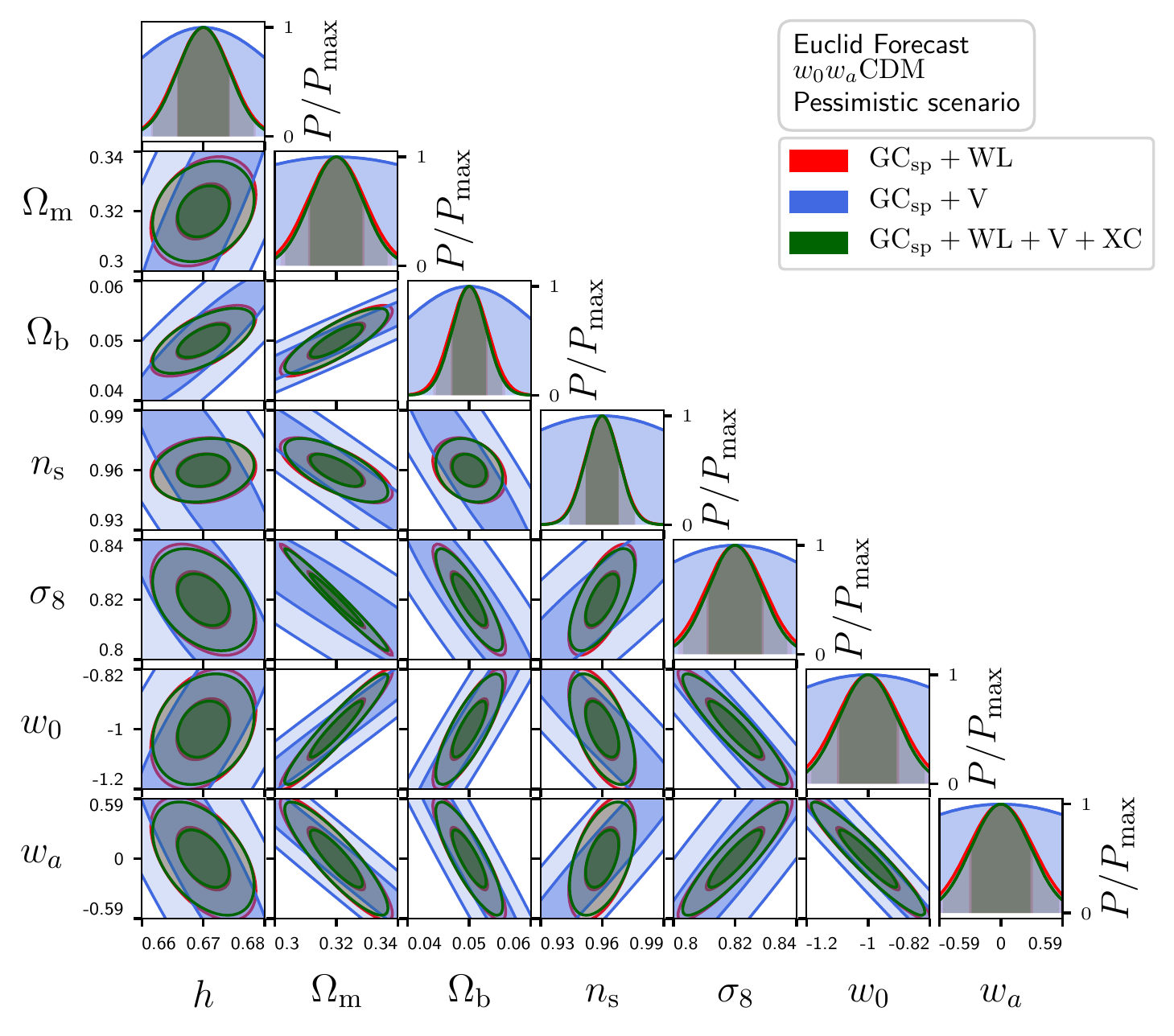}\\ \includegraphics[width=0.73\textwidth]{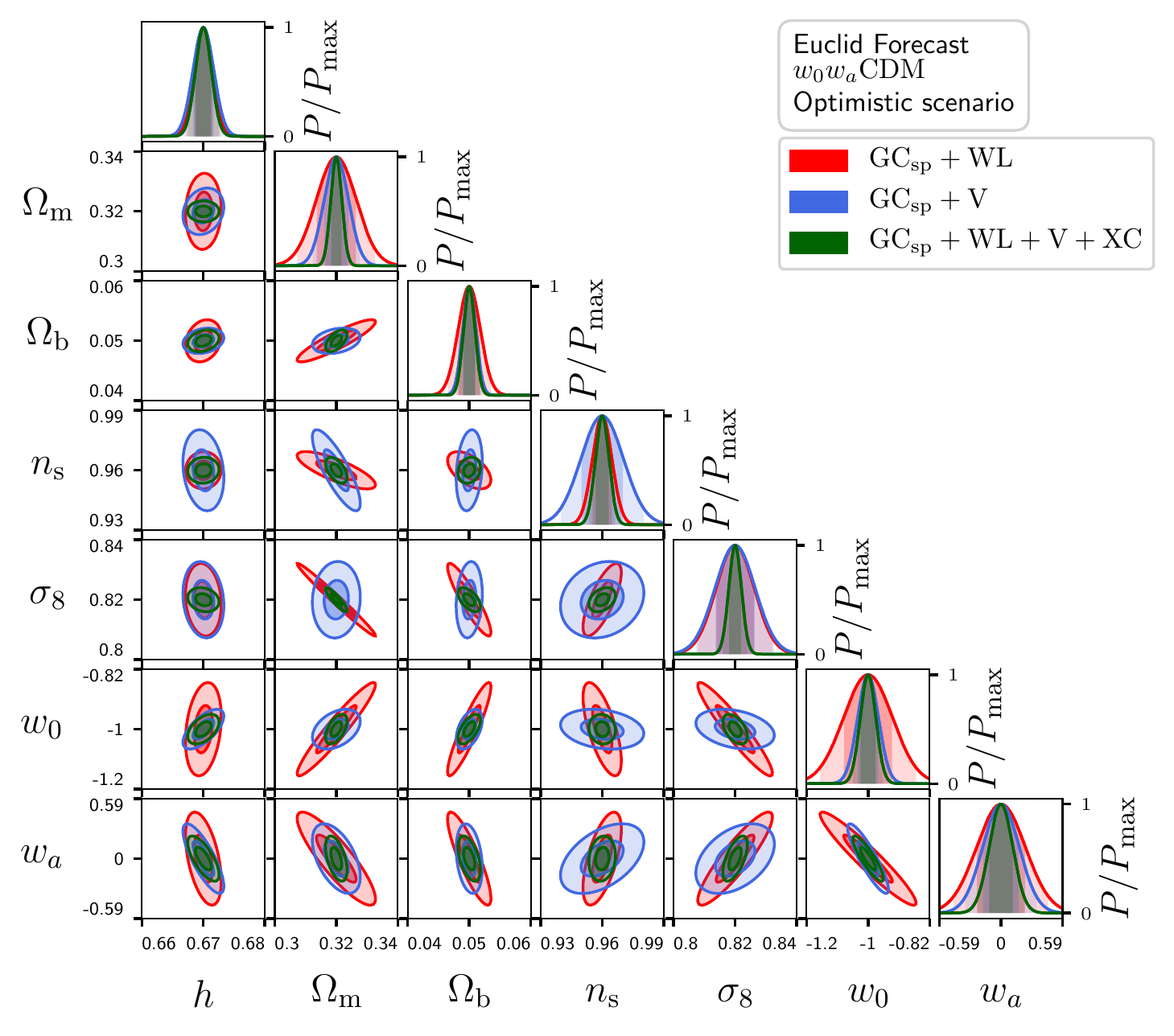}
        \end{tabular}
        \caption{Full contour plots when $\mathrm{GC_{sp}}$ is combined with $\xc$ for the $\wcdm$ model, in the pessimistic (\textit{top}) and in the optimistic (\textit{bottom}) scenarios.}
        \label{fig:GCs_r25}
        \end{figure*}

In this section we combine the spectroscopic galaxy clustering, $\text{GC}_\text{sp}$,  to the WL, void clustering, and void-lensing cross-correlation probes. To this purpose, we considered $\text{GC}_\text{sp}$ as a probe independent of the others, and made use of the $\text{GC}_\text{sp}$ Fisher matrix as provided by~\citetalias{EuclidCollaboration2019} for the observed anisotropic galaxy power spectrum, adding it to the ones computed in this work. Here our goal was to evaluate if and by how much the void clustering and the void-lensing correlation can still improve the $\Euclid$ performance even when its two primary probes are both accounted for. The corresponding results are reported in Fig.~\ref{fig:GCs_r25} and summarized in Table~\ref{tab:FoM_marginalised_errors_gcsp}.
\begin{table*}[htbp]
\caption{Marginalized 1-$\sigma$ errors in different cosmological scenarios for WL+V+$\mathrm{GC}_\mathrm{sp}$, when they are assumed to be independent, together with their combinations when the void-lensing cross-correlation is included. In each probe block, the first row shows the errors for the pessimistic scenario, while the second row corresponds to the optimistic one.}
\centering
\setlength{\tabcolsep}{2pt}
\begin{tabular}{l |c |c |c |c |c |c |c |c | c}
\toprule
Probe  
& $h$ & $\Omm$ & $\Omb$ & $\sige$ & $\ns$ & $\Mnu[\mathrm{eV}]$ & $\wz$ & $\wa$ & FoM\\
\midrule
\multicolumn{10}{c}{{$w_0w_a$CDM}}\\
\midrule
\multirow{2}{*}{$\wlpv+\mathrm{GC}_{\mathrm{sp}}$}
& 0.00435 & 0.00931 & 0.00301 & 0.00947 &  0.00832 & -- & 0.0923 & 0.31 & 113\\
& 0.00133 & 0.00395 & 0.00128 & 0.0065  &  0.0036 & -- &  0.0256 & 0.196 & 702\\
\hline
\multirow{2}{*}{$\xc+\mathrm{GC}_{\mathrm{sp}}$}
& 0.00420 & 0.00856 & 0.00295 & 0.0092  &  0.00811  & -- & 0.0883 & 0.295 & 124\\
& 0.0013  &  0.0018 & 0.00088 & 0.00205 &  0.00332  & -- & 0.0228 & 0.113 & 791\\
\bottomrule
\end{tabular}
\label{tab:FoM_marginalised_errors_gcsp}
\end{table*}

In the pessimistic scenario, constraints given by $\text{GC}_\text{sp}+\text{V}$ are weaker than the ones given by $\text{GC}_\text{sp}+\text{WL}$; for instance, the FoM has a value of 15 in the former case, and of 112 in the latter one. We find that the constraints given by $\text{GC}_\text{sp}+\text{WL}+\text{V}$ are only slightly improved with respect to $\text{GC}_\text{sp}+\text{WL}$. Nevertheless, the constraints given by $\text{GC}_\text{sp}+\text{WL}+\text{V}+\text{XC}$ are tighter than in the $\text{GC}_\text{sp}+\text{WL}$ case. In particular, the FoM increases by $\sim5\%$ when considering $\text{GC}_\text{sp}+\text{WL}+\text{V}+\text{XC}$ against $\text{GC}_\text{sp}+\text{WL}$.

In the optimistic scenario, constraints given by $\text{GC}_\text{sp}+\text{V}$ are comparable to the ones given by $\text{GC}_\text{sp}+\text{WL}$. Moreover, the FoM increases from 192 for $\text{GC}_\text{sp}+\text{WL}$, up to 702 for $\text{GC}_\text{sp}+\text{WL}+\text{V}$, in which case constraints on the other cosmological parameters become tighter: by $\sim 10\%$ for $h$, by $\sim 30\%$ for $\ns$, by $\sim50-55\%$ for $\Omb$ and $\sige$, and by $\sim 70\%$ for $\Omm$.

Finally, when also including the void-lensing cross-correlation, while in the pessimistic case there is at most a 10\% improvement in the FoM, in the optimistic setup we find that the most improved parameters are $\Omm$ and $\sige$, with their constraints improved by a factor of 2 and 3, respectively; the total FoM increases by $\sim 10\%$, that is, from 702 for $\text{GC}_\text{sp}+\text{WL}+\text{V}$, up to 791 for $\text{GC}_\text{sp}+\text{WL}+\text{V}+\text{XC}$.

\begin{figure*}[tbp]
\centering
\hspace*{-0.83cm}
        \begin{tabular}{l l}
             \includegraphics[width=0.73\textwidth]{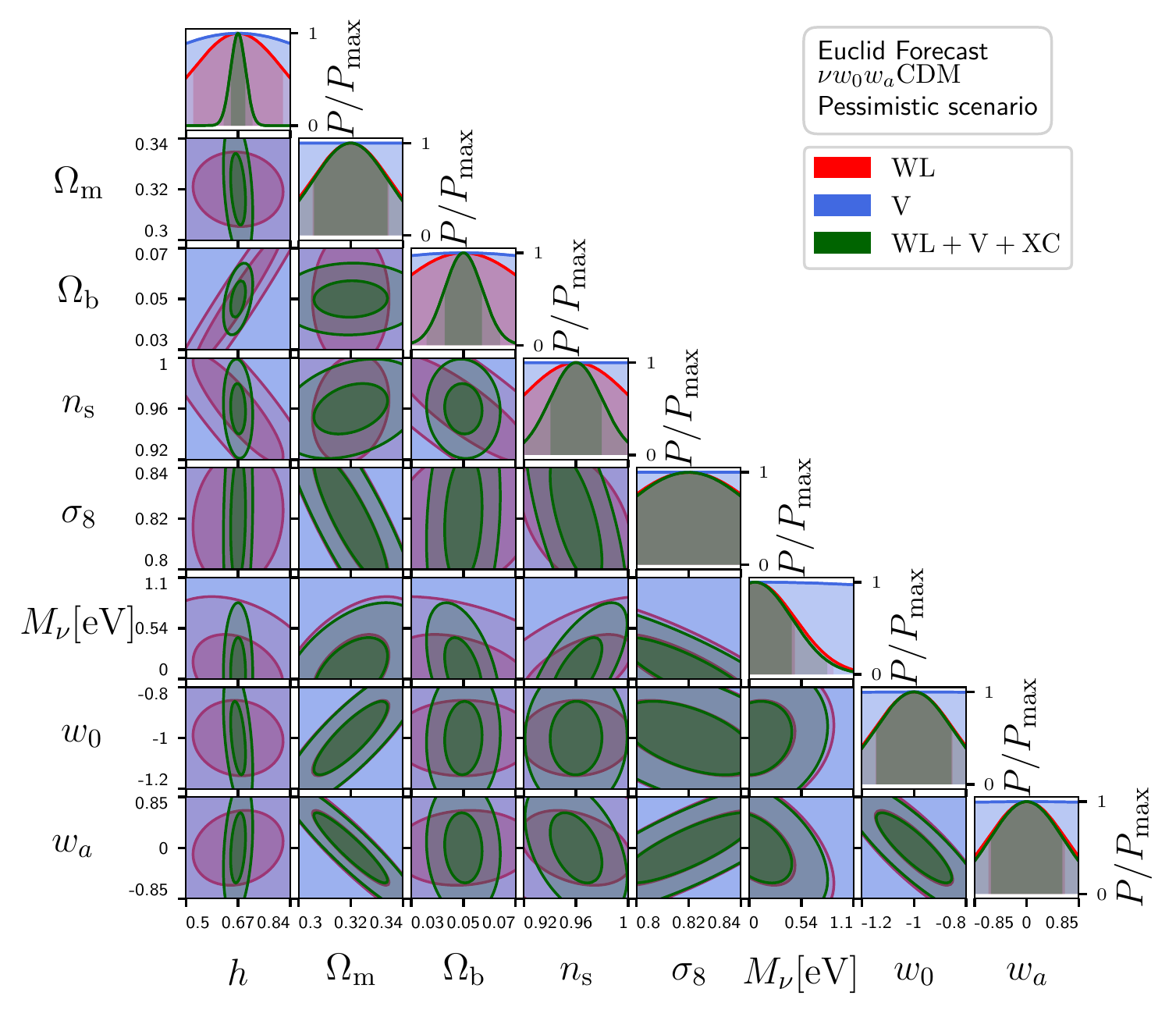}\\ \includegraphics[width=0.73\textwidth]{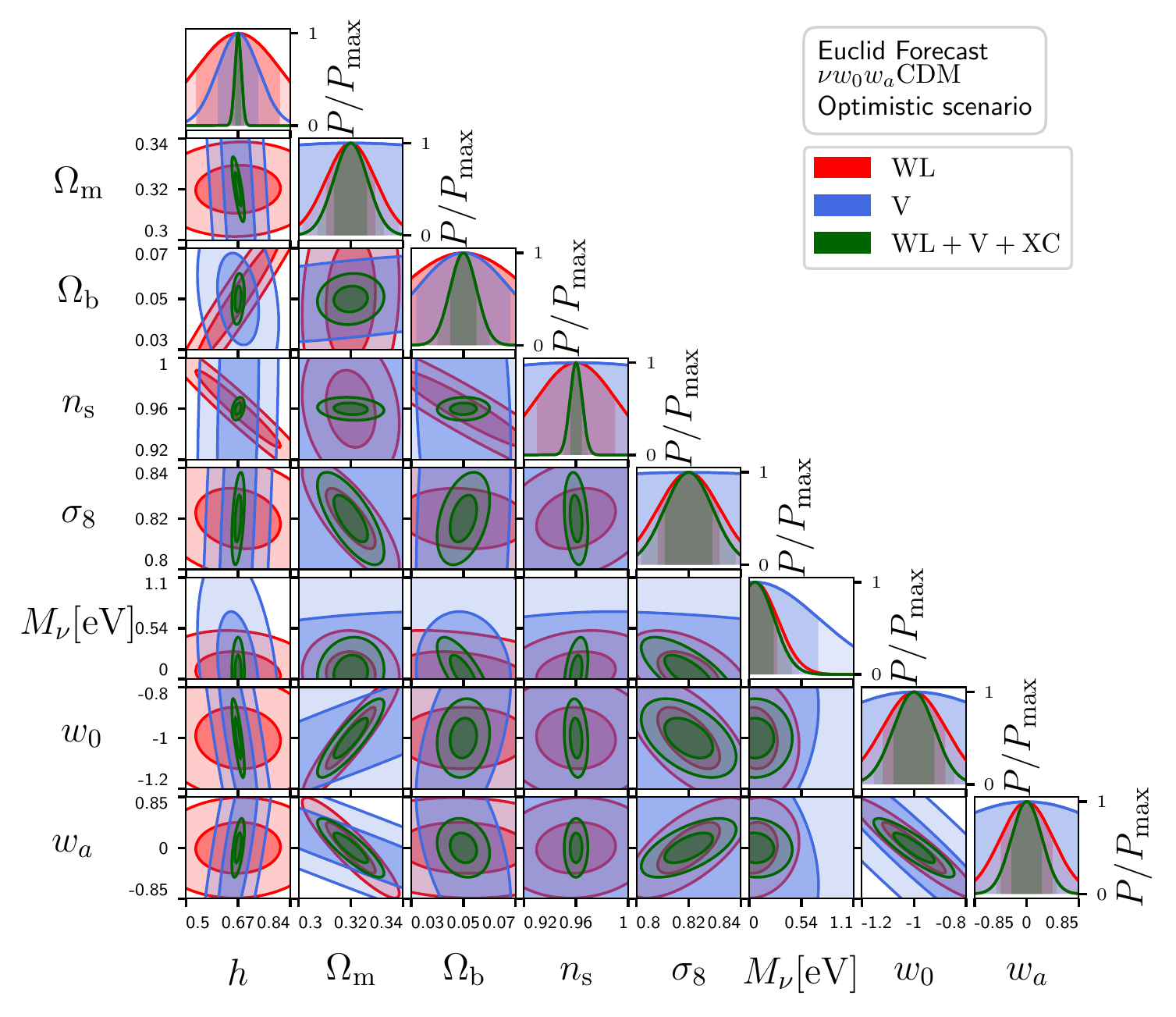}
        \end{tabular}
        \caption{Fisher matrix marginalized contours for the $\nwcdm$ model, in the pessimistic (\textit{top}) and in the optimistic (\textit{bottom}) scenarios.}
        \label{fig:nwcdm_pess_opt}
        \end{figure*}

\subsection{Combining massive neutrinos with dynamical dark energy}

In this section we consider $\Mnu$, $\wz$, and $\wa$ as free parameters. The contour plots for this model cosmology are reported in Fig.~\ref{fig:nwcdm_pess_opt}, both for the pessimistic and optimistic cases. As expected, in this scenario the constraint on $\sige$ gets even weaker with respect to the baseline $\Lambda$CDM case, due to the combined variation of $\Mnu$, $\wz$, and $\wa$, entering the matter power spectrum, the Hubble parameter, and the growth factor. Comparing the pessimistic and optimistic scenarios, we observe improvements on the constraints in the latter case, but the orientation of the ellipses remains unchanged: this is reassuring as it means that the results do not change qualitatively when the two different setups are considered. In addition, we also performed a forecast for this model with a fiducial neutrino mass of $\Mnu=0.33$ eV with a degenerate mass spectrum, using the best-fit value in \cite{Pellejero-Ibanez2016}: with this choice, the marginalized errors on $\Mnu$ decrease by $\sim 10\%$ with respect to the fiducial case with $\Mnu=\SI{0.06}{eV}$. Again, this can be explained since, by increasing the value of $\Mnu$, we increase the impact of neutrinos on the matter power spectrum, which suffers a larger suppression with respect to the fiducial neutrino mass case, becoming more sensitive to the presence of neutrinos. 

From Table~\ref{tab:FoM_marginalised_errors_all}, it is possible to observe that, when we consider the case including the void-lensing cross-correlation signal compared to the one where the two probes are joined independently, the error on $\Mnu$ gets reduced by  $\sim5\%$ and $\sim10\%$, and the FoM is enhanced by $10\%$ and $25\%$, in the pessimistic and optimistic scenarios, respectively.

\begin{figure*}[tbp]
\hspace*{-0.25cm}
        \begin{tabular}{c@{}c}
             \includegraphics[width=0.49\textwidth]{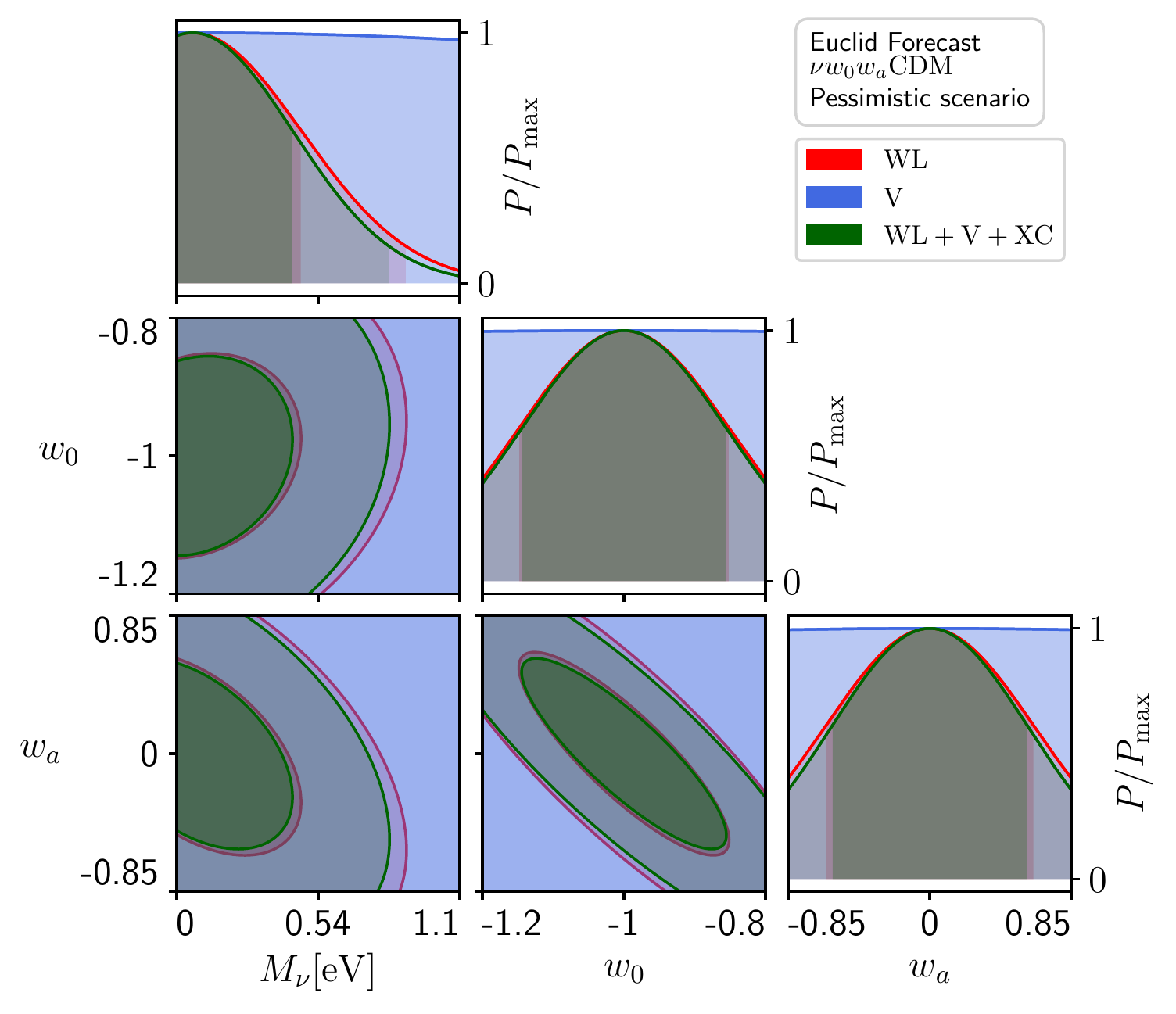}&  \includegraphics[width=0.49\textwidth]{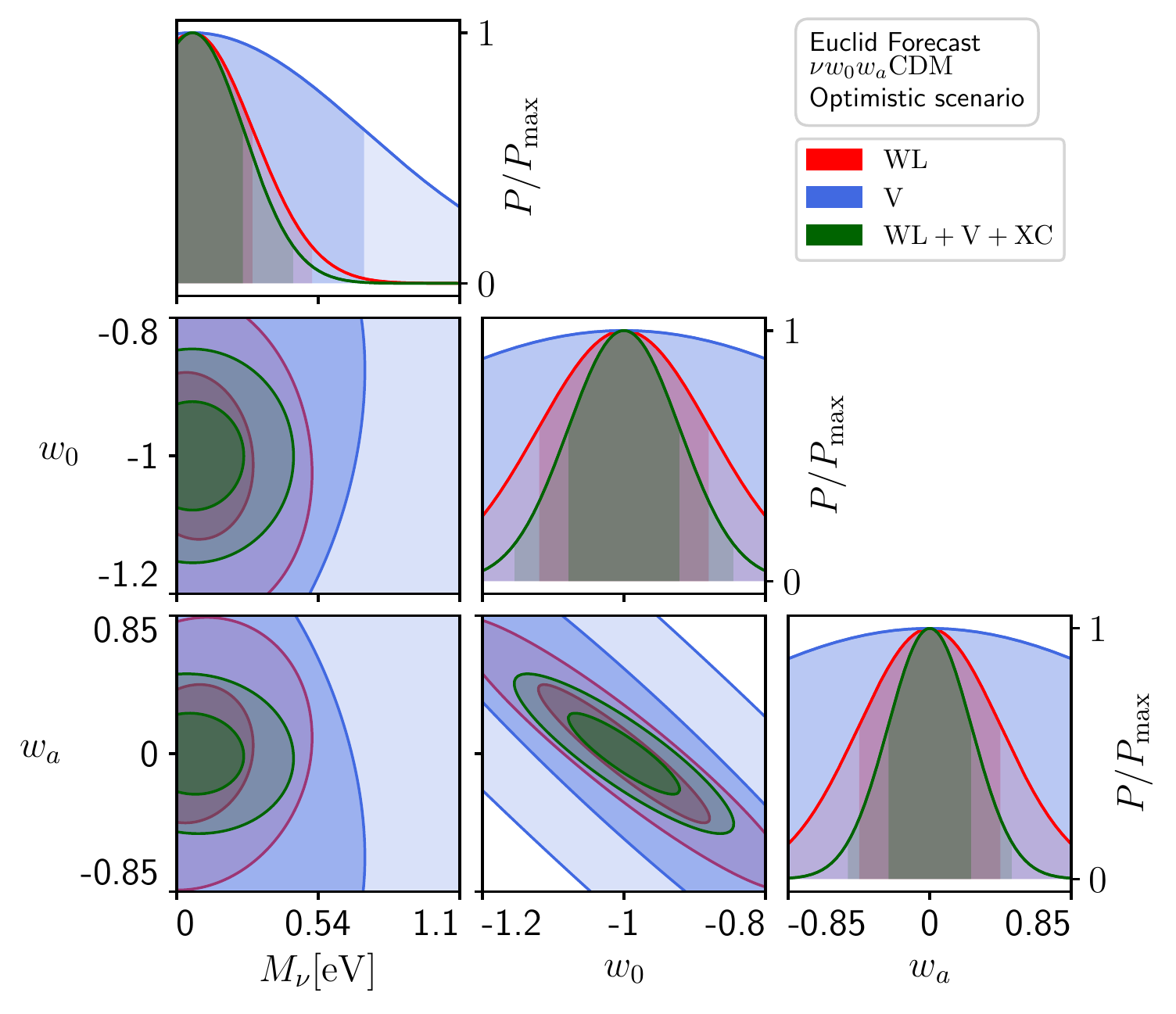}
        \end{tabular}
        \caption{Forecast contours for the $\nwcdm$ model, in the pessimistic (LEFT) and in the optimistic (RIGHT) scenarios.}
        \label{fig:nwcdm_interm_both_r}
        \end{figure*}

As a final consideration, we comment on the mutual impact of $\Mnu$ and DE. Including $\wz$ and $\wa$ increases the error on $\Mnu$ and vice versa. The increase in the $\wz$ and $\wa$ errors, due to the variation of $\Mnu$, is at most $\sim15\%$, while the error on $\Mnu$ increases at most by $70\%$ due to the variation of $\wz$ and $\wa$. However, especially for $\wz$ and $\wa$, the constraints are less impacted when the void-lensing cross-correlation is included in the analysis, compared to the constraints given by the single probes alone, as shown in the top panels of Fig.~\ref{fig:nwcdm_interm_both_r}.

\subsection{Systematics checks}
\label{sec:syst}

The analysis presented in this work relies on various choices for the computation of the Fisher matrices, such as the number of bins in multipoles $\ell$, redshift $z$, and wave-number $k$, as well as on differentiation methods.
In this section we present the stability of the final results against different computation choices.

We performed stability tests where the multipole range was divided into 80, 100, and 120 logarithmically equispaced bins, where the $\cl$ were evaluated in the linear center of the bin. The impact of this $\ell$-binning was negligible: the marginalized 1-$\sigma$ errors stayed unchanged up to the third digit.
We also performed a forecast with a reduced multipole range, from $\ell=20$ to $\ell=1350$. Both for the pessimistic and optimistic void bias scenarios, the impact of the multipole range reduction on the FoM and $\Mnu$ 1-$\sigma$ uncertainty was about $10\%$.

The impact of the $k$-binning was negligible too.
$\camb$ evaluates the matter power spectrum $\pmkz$ with a logarithmic $k$-binning grid. For the reported forecasts, $\pmkz$ was evaluated with 60 $k$ values per decade.
Stability tests were performed also using 50 and 90 $k$ values per decade; the marginalized 1-$\sigma$ errors remained unchanged up to the third digit. 

The impact of the $z$-binning was minor.
$\camb$ evaluates the matter power spectrum $\pmkz$ with a linear $z$-binning grid.
For the reported forecast, $\pmkz$ was evaluated for 450 $z$ values.
Stability tests were also performed using 300 and 600 $z$-bins: the differences in the marginalized 1-$\sigma$ errors were always smaller than $3\%$.

The impact of the differentiation method was also negligible. In the reported forecast, the SteM derivative method was used. A forecast was also performed with the semi-analytical derivative method; the marginalized 1-$\sigma$ errors stayed unchanged up to the fourth digit.

We also performed a forecast using equi-populated redshift bins for voids, rather than the equi-populated redshift bins for galaxies. This modification led to slightly (about $10$-$15\%$) worse constraints. This can be understood looking at the void distribution in Fig.~\ref{fig:voids_distribution}, which increases with the redshift $z$. Thus the requirement of equi-populated redshift bins translates into broader (and so sparser) bins at low $z$, and narrower (and so denser) bins at high $z$, with respect to the equi-populated galaxy configuration.  Since the effective void bias decreases in absolute value with $z$, this leads to more bins with a lower void bias, resulting in a decrease in the $\cl$ in those bins, and hence to weaker constraints, with, as expected, a larger impact in the optimistic case than in the pessimistic case, where we marginalize over the void bias.
Overall we conclude that our results are robust against various computation choices.

\section{Conclusion and outlook}
\label{sec:conclusion}
In this work we present the first forecast on cosmological parameter inference obtained combining the void-lensing cross-correlation with the WL and void angular two-point correlations, as will be measured from the \Euclid photometric galaxy catalogs. In order to reach this goal we considered the following approach:

\begin{itemize}
    \item we measured the projected void density distribution in redshift, $n^\mathrm{v}(z)$, directly from the \Euclid Flagship photometric mock galaxy catalog, using the 2D void finder of~\cite{Sanchez2017};

    \item we evaluated the $\cl$ numerical derivatives, which enter the Fisher matrix expression, using the SteM technique \citep{Camera2017}. We tested the stability of results against different differentiation techniques, and found them to be robust;

    \item we used the void bias as obtained from the PBS and excursion set formalisms~\citep{Sheth2004}, and considered the volume-conserving $V\,{\rm d}n$ void size function model~\citep{Jennings2013}.
\end{itemize}

We present our results for two different choices of the effective void bias $\bveff(z)$: an ``optimistic scenario'', where we exploited the cosmology dependence of both the void size function and void bias,
and a ``pessimistic scenario'', where we assumed the void bias evolution to be given by the growth factor in the reference cosmology, but its absolute normalization was supposed to be unknown, so that $\bveff(z=0)$ was marginalized over as a nuisance parameter.

We present parameter forecasts for different cosmological models: starting from a flat $\lcdm$ model, we first separately added as free parameters the total neutrino mass and the CPL parametrization of the DE equation of state~\citep{chevallier2001accelerating, Linder2002}, then we add both in combination. Our main findings are presented in Table~\ref{tab:FoM_marginalised_errors_all} and Figs.~\ref{fig:lcdm_pess_opt}--\ref{fig:nwcdm_pess_opt}, and can be summarized as follows:
\begin{itemize}
\item[$\bullet$] {\it \textbf{$\lcdm$ cosmology}}: the WL angular correlation function is able to constrain all the cosmological parameters better than the photometric void angular spectrum; however, void constraints on $h$ and $\Omb$ are competitive in the optimistic scenario. For the Hubble constant, $h$, this is due to the form of the weak lensing kernel and the integration along the line of sight. For $\Omb$ this is due to the presence of the BAO features in the void angular spectrum, which on the other hand are washed out in the lensing angular spectrum (see, e.g., Fig.~\ref{fig:cl_plots}). In general, these, together with $\ns$, are the parameters for which the constraints improve most when void clustering is combined with WL.\\

\item[$\bullet$] {\it \textbf{$\nlcdm$ cosmology}}: adding to the analysis the total neutrino mass, $\Mnu$, as a free parameter mainly impacts the constraints on $\sige$, as expected, since $\sige$ and $\Mnu$ both affect the amplitude of the matter power spectrum (even if in the $\Mnu$-case this is a scale-dependent effect). The void clustering, together with the void-lensing cross-correlation, improves the determination of the neutrino mass scale with respect to WL alone: the constraint on $\Mnu$ becomes stronger by $\sim5\%$ and $\sim15\%$, respectively, in the pessimistic and optimistic scenarios.\\

\item[$\bullet$] {\it \textbf{$\wcdm$ cosmology}}: when the CPL parametrization of the DE equation of state is considered, we find that the most affected parameters are $h$, $\Omm$, and $\sige$. The expression of the Hubble parameter Eq.~\eqref{eq:hubble} explains the effect on $h$ and $\Omm$, while the impact of $\wz$ and $\wa$ on the linear growth factor~\citep{Linder2003} accounts for the effect on $\sige$ and, again, $\Omm$. Also, in this scenario void clustering helps to tighten the constraints on the cosmological parameters. In particular, the DE FoM increases by $\sim10\%$ in the pessimistic scenario and by a factor of $\sim2$ in the optimistic scenario.\\

\item[$\bullet$] {\it \textbf{$\nwcdm$ cosmology}}: in this model we evaluated, in particular, the mutual impact among $\wz$, $\wa$, and $\Mnu$. On the one hand, we find that, when adding $\Mnu$, the marginalized errors of $\wz$ and $\wa$ increase at most by $\sim$ $15\%$. On the other hand, when adding $\wz$, $\wa$ as free parameters, the impact on the $\Mnu$ constraints is higher, as its error increases even by $70\%$.
However, especially for $\wz$ and $\wa$, the constraints are less impacted when the void-lensing cross-correlation is included, compared to the constraints provided by the two probes combined independently.\\

\item[$\bullet$] {\it \textbf{FoM and the void-lensing cross-signal}}: in order to evaluate the improvement in the constraining power provided by adding the void-lensing cross-correlation signal to void clustering and weak lensing, we also considered the case where these two probes were assumed to be independent.
When including the void-lensing correlation, the error on $\Mnu$ was reduced by $5\%$ and $10\%$, in the pessimistic and optimistic scenarios, respectively. The FoM was enhanced by $10\%$ and $25\%$, in the pessimistic and optimistic scenarios, respectively.\\

\item[$\bullet$] {\it \textbf{Adding ${\rm GC_{sp}}$}}: finally, we combined the galaxy lensing, void clustering, and void-lensing cross-correlation probes with the spectroscopic galaxy clustering, ${\rm GC}_{\rm{sp}}$. In this case, we exploited the Fisher matrices as obtained by~\citetalias{EuclidCollaboration2019} in the so-called ``pessimistic'' and ``optimistic'' settings, and considered the flat $\wcdm$ scenario, since ${\rm GC}_{\rm{sp}}$ forecasts are computed keeping $\Mnu=0.06 \, {\rm eV}$ fixed. When combining $\xc$ with ${\rm GC}_{\rm{sp}}$, both assumed in the pessimistic configuration, we find that the FoM increases from $23$ for $\xc$ alone up to 117 for $\xc$ + ${\rm GC}_{\rm{sp}}$, and from $105$ up to $791$ in the optimistic case. In particular in the latter case, results are promising and competitive with other kinds of probe combination, as, for example, when galaxy lensing is combined with photometric galaxy clustering, ${\rm GC}_{\rm ph}$,  and their cross-correlation~\citep{Tutusaus2020}. 
As intrinsic alignments are not included in this analysis, we cannot make a proper comparison with the results reported in Table~14 in~\citetalias{EuclidCollaboration2019}, where the authors show a total FoM of $377$ and $1257$ in the pessimistic and optimistic settings, respectively. The results are, nevertheless, extremely encouraging: we show that including void clustering and the void-lensing cross-correlation in the total likelihood analysis will allow us to considerably improve $\Euclid$'s performance.
\end{itemize}

The forecasts presented in this work show that photometric void clustering and its cross-correlation with WL deserve to be exploited in the data analysis of the \Euclid galaxy survey, as they could be able to improve constraints on several cosmological parameters, such as, in particular, the total neutrino mass~\citep{Kreisch2021} and the DE equation of state.
In this respect, it is worth noting that, for a full comparison with the Euclid performance from primary probes, in our analysis we should have included also the information from photometric galaxy clustering. However, we decided not to add this probe to the analysis in order to avoid double counting the information from the photometric sample. A way to include $\textrm{GC}_\textrm{ph}$, without incurring in this issue, is offered by a full-field inference approach, which allows us to self-consistently analyze GC and voids in the DM distribution~\citep{Leclercq:2014pga}. It would be important to also include $\textrm{GC}_\textrm{ph}$ since, as shown in this work, the combination of void-clustering, void-lensing, and $\textrm{GC}_\textrm{sp}$ does not allow us to reach the same Euclid performance as the combination of \Euclid primary probes~\citepalias{EuclidCollaboration2019}. However, this does not imply that the information coming from the photometric void sample should be neglected, as voids and galaxies are tracers with a very different bias, and hence probe different regimes of LSS~\citep{Wang:2020dtd}, therefore providing therefore complementary information about our Universe.
Moreover, in this paper we only consider only some observational and astrophysical uncertainties, neglecting baryonic physics and intrinsic alignment of galaxies.

While the aim of this work was to show that the inclusion of the void-void auto-correlation and void-lensing cross-correlation will improve the Euclid survey performance, and we performed some checks against the Flagship simulation to ensure the reliability of our theoretical modeling, additional work needs to be carried out to prepare the pipelines for the analyses of forthcoming real data.
First and foremost, the theoretical model should be further improved. As already mentioned, we did not include nonlinear scales, since this would require a modeling of the void density profile and its dependence on cosmological parameters. 
Second, we need to rely on more realistic mock data, which will account for more effects: Euclid survey specification and systematics and possibly mispecifications induced by the HOD, which could induce some errors in the identified void catalogs. Finally, the framework presented here can also be easily extended to investigate non-flat scenarios, more specific DE models, theories of modified gravity, and primordial non-Gaussianities \citep{Chan2019}.

\begin{acknowledgements}
MB acknowledges financial support from the ASI agreement n. I/023/12/0 "Euclid attivitá relativa alla fase B2/C". NH is supported by the Excellence Cluster ORIGINS, which is funded by the Deutsche Forschungsgemeinschaft (DFG, German Research Foundation) under Germany's Excellence Strategy -- EXC-2094 -- 390783311. AP is supported by NASA ROSES grant 12-EUCLID12-0004, and NASA grant 15-WFIRST15-0008 to the Nancy Grace Roman Space Telescope Science Investigation Team ``Cosmology with the High Latitude Survey''. AK has been supported by a Juan de la Cierva \emph{Incorporaci\'on} fellowship with project number IJC2018-037730-I, and funding for this project was also available in part through SEV-2015-0548 and AYA2017-89891-P. SN acknowledges support from an STFC Ernest Rutherford Fellowship, grant reference ST/T005009/1. 
\AckEC
\end{acknowledgements}

\bibliographystyle{aa} 
\bibliography{Euclid-Voids}

\appendix

\section{Fisher matrix expression proof}
\label{sec:appendix_C}
This paper presents a forecast for cosmological parameter measurements using the Fisher matrix technique. This appendix is devoted to the derivation of the Fisher matrix expression considering as observables either the $a_{\ell m}$ (field perspective) or the $\cl$ (estimator perspective).

\subsection{The field perspective}
We consider the spherical harmonics' expansion coefficients $a_{\ell m}^{\mathrm{A}i}$, the coefficients of the spherical harmonics decomposition of the 2D field $\rm A$ in the $i$-th tomographic bin, which we assume to follow a multivariate gaussian distribution with zero mean and covariance matrix $\mathbf{S}$
\begin{equation}
L(\bm{a}_{\ell}|\bm{\theta})=\frac{\exp \left(-\frac{1}{2}\bm{a}_{\ell}^{\mathrm{T}} {\mathbf{S}}^{-1}\bm{a}_{\ell}\right)}{\sqrt{(2 \pi)^{n}|{\mathbf{S}}|}}\, ,
\end{equation}
where $\bm\theta$ is the vector of the parameters and $\bm{a}_{\ell}$ is a vector collecting the $a_{\ell m}^{\mathrm{A}i}$. The logarithm of this probability distribution function is
\begin{align}
\ln L(\bm{a}_{\ell}|\bm{\theta}) &= 
-\frac{1}{2}\left[
n\ln(2\pi) + \ln |\mathbf{S}| + \bm{a}_{\ell}^{\mathrm{T}} {\mathbf{S}}^{-1}\bm{a}_{\ell}
\right]
\nonumber\\
&=-\frac{1}{2}\left[
n\ln(2\pi) + \ln |\mathbf{S}| +\operatorname{Tr}(\mathbf{S}^{-1}\mathbf{A})
\right]\, ,
\label{eq:loglikelihood}
\end{align}
with $\mathbf{A} \equiv \bm{a}_{\ell} \bm{a}_{\ell}^T$. The computation of the second derivative, after taking the expectation value, yields
\begin{equation}\label{eq:fisher_field_first_expr}
F_{\alpha\beta}(\ell) = 
-\left\langle
\ln L(\bm{a}_{\ell}|\bm{\theta})_{,\,\alpha\beta}
\right\rangle
=\frac{1}{2}\operatorname{Tr}
\left(
\mathbf{S}^{-1}\mathbf{S}_{,\,\alpha}\mathbf{S}^{-1}\mathbf{S}_{,\,\beta}
\right)\, ,
\end{equation}
where $\mathbf{S}=\left\langle \mathbf{A}\right\rangle$ and we used the following matrix identities:
\begin{align}
(\ln |\mathbf{S}|)_{,\,\alpha} = \operatorname{Tr}(\mathbf{S}^{-1}\mathbf{S}_{,\,\alpha})\, , 
\qquad \qquad
(\mathbf{S}^{-1})_{,\,\alpha} = -\mathbf{S}^{-1}\mathbf{S}_{,\,\alpha}\mathbf{S}^{-1}\, .
\end{align}
Now, we specify how the $a_{\ell m}^{\mathrm{A}i}$ were collected into the vector $\bm{a}_{\ell}$. First of all, we recall that
\begin{equation}\label{eq:alm_prod_expect_val}
\left\langle  a_{\ell m}^{\mathrm{A}i}a_{\ell' m'}^{\mathrm{B}j}  \right\rangle= \Sigma_{ij}^{\mathrm{AB}}(\ell)\delta_{\ell\ell'}\delta_{mm'}\,,  \qquad 
\Sigma_{ij}^{\mathrm{AB}}(\ell) = \clABij+N_{ij}^{\mathrm{AB}}(\ell)\, ,
\end{equation}
where $N_{ij}^{\mathrm{AB}}$ is the Poisson shot noise for probes combination $\mathrm{AB}$ in tomographic bins $ij$.
Working at fixed multipole $\ell$, the vector $\bm{a}_{\ell}$ has a multi-index $I = (m, i, \mathrm{A})$. Each of these indices runs in a different range, and the range of the tomographic index $i$ may depend on the probe $A$ considered. To summarize:
\begin{itemize}
    \item $m$ varies from $-\ell$ to $\ell$
    \item $i$ varies from 1 to $N_\mathrm{A}$
    \item $\mathrm{A}$ varies from 1 to $\mathcal{N}$,
\end{itemize}
with $\mathcal{N}$ being the number of probes and $N_\mathrm{A}$ the number of tomographic bins for probe $\mathrm{A}$.
We chose to order the array $\bm{a}_\ell$ varying the three indices with a significance increasing from left to right. With this choice the matrix $\mathbf{S}$ has the following block form:
\begin{align}\label{eq:kronecker_product_covariance}
\mathbf{S} = {\Sigma} (\ell) \otimes \mathbb{1}_{2\ell+1},
\qquad
{\Sigma} (\ell) = 
\begin{pmatrix}
{\Sigma}^{11} (\ell)  & 
{\Sigma}^{12} (\ell)  & 
\cdots & 
{\Sigma}^{1\mathcal{N}} (\ell)  \\[1ex]
{\Sigma}^{21} (\ell)  & 
{\Sigma}^{22} (\ell)  & 
\cdots & 
{\Sigma}^{2\mathcal{N}} (\ell)  \\[1ex]
\vdots          & \vdots          & \ddots & \vdots \\[1ex]
{\Sigma}^{\mathcal{N}1} (\ell)  & {\Sigma}^{\mathcal{N}2} (\ell)  & 
\cdots & {\Sigma}^{\mathcal{N}\mathcal{N}} (\ell) 
\end{pmatrix}.
\end{align}
Here ${\Sigma}^{\mathrm{AB}}(\ell)$ is the tomographic covariance matrix, that is to say,$[{\Sigma}^{\mathrm{AB}}(\ell)]_{ij} = \Sigma^{\mathrm{AB}}_{ij}(\ell)$,
and $\otimes$ is the Kronecker product. Equation \eqref{eq:kronecker_product_covariance} can be understood as follows: the matrix $S$ is made up of diagonal blocks of size $\ell$, each of them proportional to the identity $\mathbb{1}_{2\ell+1}$ with a different factor $\Sigma^{\mathrm{AB}}_{ij}(\ell)$. Here the identity matrix $\mathbb{1}_{2\ell+1}$ is exactly the Kronecker delta $\delta_{mm^{\prime}}$, which appears in Eq.~\eqref{eq:alm_prod_expect_val}. In order to compute the trace in Eq.~\eqref{eq:fisher_field_first_expr} we make use of some properties of the Kronecker product \citep{SteebKronProd}:
\begin{align}\label{eq:kron_prod_properties}
&(\mathbf{A} \otimes \mathbf{B}) (\mathbf{C} \otimes \mathbf{D})
= \mathbf{A}\mathbf{C} \otimes \mathbf{B}\mathbf{D},\\
&(\mathbf{A} \otimes \mathbf{B})^{-1}= \mathbf{A}^{-1} \otimes \mathbf{B}^{-1}\, ,\\
&\operatorname{Tr}(\mathbf{A} \otimes \mathbf{B}) 
=\operatorname{Tr}(\mathbf{A}) \operatorname{Tr}(\mathbf{B})\, .
\end{align}
Using these properties, we have
\begin{align*}
\operatorname{Tr} 
\left[ 
\mathbf{S}^{-1} \mathbf{S}_{,\,\alpha}
\mathbf{S}^{-1} \mathbf{S}_{,\,\beta} 
\right]
&= \operatorname{Tr} \left[\left(
\mathbf{\Sigma}(\ell)^{-1} 
\mathbf{\Sigma}(\ell)_{,\,\alpha}
\mathbf{\Sigma}(\ell)^{-1}
\mathbf{\Sigma}(\ell)_{,\,\beta}
\right) \otimes \mathbb{1}_{2\ell+1} 
\right] \\[0.5ex]
&= \operatorname{Tr}
\left(
\mathbf{\Sigma}(\ell)^{-1} 
\mathbf{\Sigma}(\ell)_{,\,\alpha}
\mathbf{\Sigma}(\ell)^{-1}
\mathbf{\Sigma}(\ell)_{,\,\beta}
\right)
\operatorname{Tr} \left( \mathbb{1}_{2\ell+1} \right) \\[0.5ex]
&= (2\ell+1) \operatorname{Tr} 
\left(
\mathbf{\Sigma}(\ell)^{-1} 
\mathbf{\Sigma}(\ell)_{,\,\alpha}
\mathbf{\Sigma}(\ell)^{-1}
\mathbf{\Sigma}(\ell)_{,\,\beta}
\right)\\[0.5ex]
&= (2\ell+1) \operatorname{Tr} 
\left(
\mathbf{\Sigma}(\ell)^{-1} 
\mathbf{C}(\ell)_{,\,\alpha}
\mathbf{\Sigma}(\ell)^{-1}
\mathbf{C}(\ell)_{,\,\beta}
\right)\, ,
\end{align*}
where in the last line we used the fact that the shot noise is independent of the cosmological parameters, and the $\cl$'s block matrix $\mathbf{C}(\ell)$ has the same form of $\mathbf{\Sigma}(\ell)$ of Eq.~\eqref{eq:kronecker_product_covariance}, with entries $\clABij$ instead of $\Sigma(\ell)^{\mathrm{AB}}_{ij}$. Then from Eq.~\eqref{eq:fisher_field_first_expr}, we can finally write
\begin{equation*}
\fisher(\ell)=\frac{2\ell+1}{2}
\operatorname{Tr}
\left(
\mathbf{\Sigma}(\ell)^{-1} 
\mathbf{C}(\ell)_{,\,\alpha}
\mathbf{\Sigma}(\ell)^{-1}
\mathbf{C}(\ell)_{,\,\beta}
\right)\, .
\end{equation*}
Now, we sum over the multipoles $\ell$, considering them as independent
\begin{equation}\label{eq:fisher_field}
\fisher = 
\sum_{\ell}\frac{2\ell+1}{2}
\operatorname{Tr}
\left[
\mathbf{\Sigma}(\ell)^{-1} 
\mathbf{C}(\ell)_{,\,\alpha}
\mathbf{\Sigma}(\ell)^{-1}
\mathbf{C}(\ell)_{,\,\beta}
\right]\, .
\end{equation}
In order to obtain the expression Eq.~\eqref{eq:fisher_analytical}, it sufficient to redefine the covariance as
\begin{align}
\mathbf{\Sigma}(\ell) 
\longrightarrow 
\sqrt{\frac{2}{(2\ell+1)\Delta\ell f_{\mathrm{sky}}}}\,
\mathbf{\Sigma}(\ell)\, .
\end{align}
In this way one accounts for a possible unequal spacing between multipole bins, weighting each term of the sum with the bin width, and with the sky fraction $f_{\mathrm{sky}}$ covered by the survey.
\subsection{The estimator perspective}
In the previous section, we evaluated the expression of the Fisher matrix for a multivariate normal distribution; then, we specialized this expression for the field perspective, when the observables are the $a_{\ell m}$'s. In this section we derive the Fisher matrix expression when the observables are the estimator $\clest$, defined as
\begin{equation}
\clestABij\equiv\frac{1}{2\ell+1}\sum_{m=-\ell}^{\ell}a_{\ell m}^{\mathrm{A} i}a_{\ell m}^{\mathrm{B} j}.
\end{equation}
One may argue that, since the $a_{\ell m}$ are normally distributed, the $\clest$ are also normally distributed. However, this ansatz is wrong; in the proceeding equation, we obtain the $\clest$ posterior distribution function and then their Fisher matrix.
Since coefficients with different $\ell$ are uncorrelated, under the Gaussian assumption, we work at a fixed $\ell$.
The $\clest$ likelihood is
\begin{equation}
L(\hat{\bm{C}}|\bm{\theta})=\int \mathrm{d}\bm{a}_{\ell}\, L(\bm{a}_{\ell}|\bm{\theta})\prod_{\mathrm{A,B}}\prod_{i,j}\delta_{\rm{D}}\left(\clestABij-\sum_{m=-\ell}^{\ell}\frac{a_{\ell m}^{\mathrm{A} i}a_{\ell m}^{\mathrm{B} j}}{2\ell+1}\right)\, .
\end{equation}
Using the Dirac delta function $\delta_\mathrm{D}$,  $L(\bm{a}_{\ell}|\bm{\theta})$ gets out from the integral
\begin{equation}
L(\hat{\bm{C}}|\bm{\theta})=\frac{f(\hat{\bm{C}})}{\sqrt{(2 \pi)^{n}|{\mathbf{S}}|}}\exp 
\left(
-\frac{1}{2}\operatorname{Tr}(\mathbf{S}^{-1}\mathbf{A})
\right)\, ,
\end{equation}
where $f(\hat{\bm{C}})$ is the result of the integral in the $a_{\ell m}$ space\footnote{Since it does not carry information on the parameters $\bm\theta$, we do not derive it explicitly. If computing the integral, one would have found the Wishart probability distribution function.}.
Taking the logarithm of the previous equation we obtain
\begin{equation}
\ln L(\hat{\bm{C}}|\bm{\theta})=\ln\left(\frac{f(\hat{\bm{C}})}{\sqrt{(2 \pi)^{n}}}\right)-\frac{1}{2}\ln(|\mathbf{S}|)-\frac{1}{2}\operatorname{Tr}(\mathbf{S}^{-1}\mathbf{A})\, .
\label{eq:loglikelihood_estimator}
\end{equation}
Proceeding as in the previous section and using $\langle \mathbf{A} \rangle = \mathbf{S}$ the expectation value of the second derivative becomes
\begin{equation}
\left\langle\ln L(\hat{\bm{C}}|\bm{\theta})_{,\,\alpha\beta}\right\rangle=
-\frac{1}{2}\operatorname{Tr}(\mathbf{S}^{-1}\mathbf{S}_{,\,\beta}\mathbf{S}^{-1}\mathbf{S}_{,\,\alpha})\, .
\end{equation}
Performing the same manipulations of the previous section, we finally obtain Eq.~\eqref{eq:fisher_field}, showing that the Fisher matrix in the field and estimator perspective is the same.

\end{document}